# Theoretical and practical aspects of the design and production of synthetic holograms for transmission electron microscopy

Paolo Rosi[1], Federico Venturi[1,2], Giacomo Medici[1], Claudia Menozzi[1], Gian Carlo Gazzadi[3], Enzo Rotunno[3], Stefano Frabboni[1,3], Roberto Balboni[4], Mohammadreza Rezaee[5], Amir H. Tavabi[6], Rafal E. Dunin-Borkowski[6], Ebrahim Karimi[5] and Vincenzo Grillo[3]

1. FIM Department, University of Modena and Reggio Emilia, 41125 Modena, Italy
2. Faculty of Engineering, University of Nottingham, Nottingham NG7 2RD, United Kingdom
3. CNR-Nanoscience Institute, S3 center, 41125 Modena, Italy
4. CNR-Institute for Microelectronics and Microsystems, 40129 Bologna, Italy
5. Department of Physics, University of Ottawa, Ottawa, Ontario K1N 6N5, Canada
6. Ernst Ruska-Centre for Microscopy and Spectroscopy with Electrons and Peter Grünberg Institute, Forschungszentrum Jülich, 52425 Jülich, Germany

**Abstract**

Beam shaping - the ability to engineer the phase and the amplitude of massive and massless particles - has long interested scientists working on communication, imaging and the foundations of quantum mechanics. In light optics, the shaping of electromagnetic waves (photons) can be achieved using techniques that include, but are not limited to, direct manipulation of the beam source (as in X-ray Free Electron Lasers (XFELs) and Synchrotrons), deformable mirrors, spatial light modulators, mode converters and holograms. The recent introduction of holographic masks for electrons provides new possibilities for electron beam shaping. Their fabrication has been made possible by advances in micrometric and nanometric device production using lithography and focused ion beam patterning. This article provides a tutorial on the generation, production and analysis of synthetic holograms for transmission electron microscopy. It begins with an introduction to synthetic holograms, outlining why they are useful for beam shaping to study material properties. It then focuses on the fabrication of the required devices from theoretical and experimental perspectives, with examples taken from both simulations and experimental results. Applications of synthetic electron holograms as aberration correctors, electron vortex generators and spatial mode sorters are then presented.

# Introduction

The transmission electron microscope was developed primarily to study matter at the highest spatial resolution. However, over time the quantum wave nature of electrons has attracted increasing interest for both fundamental reasons and applications. The wave nature of electrons provides analogies with light optics. For non-relativistic and monochromatic electrons, the Helmholtz equation can be used to describe both electrons and photons. Concepts such as the refractive index and lenses can also be considered in both contexts with similar results. In light optics, refraction results from interference of incident photons with those re-emitted by atoms in a material. For electrons, electrostatic and magnetostatic potentials result in retardation (or anticipation) of an electron wave. Analogies between the two fields have been used widely in electron microscopy. However, light optics has provided a broad range of applications beyond imaging, with recent progress (*e.g.*, superoscillation microscopy) triggered by the concept of structured light waves, whereby a wave front and its spatial intensity distribution can be controlled in a manner that goes beyond the use of conventional optical elements. Recently, the concept of structured waves has been extended to matter waves, primarily to electrons. Structured electron waves include electron beams with helical wave fronts (*i.e.*, electron vortex beams), self-accelerating beams and non-diffracting beams, as well as orbital angular momentum analyzers. One can also fabricate conventional electron optical devices such as lenses, diffractive elements and aberration correctors using a holographic approach. The key technology for electrons is the use of synthetic holograms to modulate the phase and amplitude of the electron wave. The word "hologram" comes from the Greek term for "whole writing". The ability to write both the intensity and the phase of an electron wave is achieved by the creation of an interference pattern, which is related to the relative phases of two waves.

Even though Gabor's original concept of holography was intended for electron optics [1], holograms have seen wider applications in light optics, becoming a ubiquitous concept (e.g., on Canadian dollar notes). In electron microscopy, holography normally refers to the recording of the interference of a wave perturbed by a semi-transparent specimen and a plane wave. In the present context, the recreation of a perturbed wave from a calculated interference pattern is of primary interest. For the sake of clarity, this approach is referred to as "*synthetic holography*", while a calculated pattern is referred to as a "*computer-generated hologram*". The two operations are inverse; when a fabricated interference pattern is illuminated, an electron beam that has the phase and amplitude of the original semi-transparent specimen is generated. A similar approach was historically implemented when a lack of computing resources meant that researchers could not apply numerical Fourier transforms and had to illuminate the recorded electron interference patterns with lasers to recreate images of objects.

In order to construct a synthetic hologram, one needs to scale down the equivalent of a transparent electron micrograph to the electron's scale. Unfortunately, there is no electron optical analog of a transparent object. The best approximation is given by a thin layer of a material with low electron absorption, such as carbon or silicon nitride ($Si_3N_4$). $Si_3N_4$ can be produced routinely in the form of membranes that can be inserted along the electron path. Standard nanofabrication techniques allow thickness modulations to be imprinted on lateral and depth scales of 50-100 nm. Local variations in the thickness and/or mean inner potential of the substrate then act to accelerate electrons, producing an effective refraction index variation of approximately $\Delta n/n = 10^{-5}$. Fortuitously, such a small refraction index variation over a typical modulation of 50 nm is sufficient to advance a 300 keV electron wave by exactly one wavelength. Conversely, 100 nm of Au is sufficient to damp most of the electron beam intensity. Therefore, the tools that are needed to modulate both the phase and the amplitude of an electron wave are available. Developments have proceeded historically from rough amplitude modulations of electron waves to today's fine and precise control over amplitude and phase modulations in the form of complex patterns. This tutorial provides an overview of the theoretical and numerical calculation, fabrication and analysis of synthetic electron holograms.

## Chapter 1 - Synthetic hologram formation: From calculations to computer-generated holograms

### 1.1 Theory of hologram formation

We begin by describing interference between a generic perturbed wave and a "known" reference wave to form a hologram. This approach allows us to describe both "imaging" holography and synthetic holography as a general theoretical framework. From a physical point of view, a hologram is generated by interference between a reference wave function $\Psi_{ref}(\vec{r})$ and a wave function of interest

$$\Psi_I(\vec{r}) = A_I(\vec{r}) e^{i\varphi_I(\vec{r})} \; , \tag{1}$$

where $A_I(\vec{r})$ and $\varphi_I(\vec{r})$ are the phase and amplitude of the wave function of interest, respectively. Holography involves the writing, in two dimensions, of an interference pattern between waves propagating in three-dimensional space. In vacuum, the wave equation (e.g., written in the form of the Helmholtz equation) constrains the wave behavior outside a specific two-dimensional plane. We consider a specific plane with coordinates $\vec{\rho} = (x, y)$ and an out-of-plane direction $z$. The wave function in three dimensions is

$$\Psi_{holo}(\vec{r}) = |\Psi_I(\vec{r}) + \Psi_{ref}(\vec{r})| \; , \tag{2}$$

whereas in a specific plane it is

$$\Psi_{holo}(\vec{\rho}) = |\Psi_I(\vec{\rho}) + \Psi_{ref}(\vec{\rho})| \, , \tag{3}$$

with corresponding intensity

$$I_{holo}(\vec{\rho}) = |\Psi_I(\vec{\rho})|^2 + |\Psi_{ref}(\vec{\rho})|^2 + 2\, Re[\Psi_I(\vec{\rho})\Psi^*_{ref}(\vec{\rho})] \, . \tag{4}$$

Alternatively,

$$I_{holo}(\vec{\rho}) = |\Psi_I(\vec{\rho})|^2 + |\Psi_{ref}(\vec{\rho})|^2 + 2|\Psi_I(\vec{\rho})||\Psi_{ref}(\vec{\rho})|\cos\left(\varphi_I(\vec{r}) - \varphi_{ref}(\vec{r})\right) \, , \tag{5}$$

where $\varphi_{ref}(\vec{r})$ is the phase of the reference beam. The use of a reference wave allows the phase $\varphi_I(\vec{r})$ to be made visible as an intensity modulation. It should have a known form, such as a plane wave or spherical wave (sometimes substituted by a parabolic approximation). The process is referred to as "inline" or "on-axis" holography if the waves propagate in the same direction and as "off-axis" holography if the waves propagate in different directions.

*1.2 "Image" holography for object phase reconstruction*

Imaging holography is the basis of synthetic holography. If one considers $\Psi_I(\vec{r})$ as a wave function obtained after passing a partially-electron-transparent sample with an unknown phase distribution, then holography can be used to extract this phase information.

Off-axis holography is performed by splitting a wave front into two parts, typically using a biprism. In electron microscopy, a biprism normally takes the form of a metal or metal-coated wire that has a voltage applied to it, with one part of the beam traveling through a region of interest on a specimen. The relative tilt of the two parts of the electron wave introduced by the biprism allows them to interfere with one another. The object wave interacts with the sample and gains a phase that depends on the physical features of the sample. The intensity of the hologram is given by the expression

$$\begin{aligned} I_{holo}(\vec{r}) = \Psi^2_{holo}(\vec{\rho}) &= |\Psi_I(\vec{\rho}) + \Psi_{ref}(\vec{\rho})|^2 \\ &= 1 + A_I^2(\vec{\rho}) + 2A_I(\vec{\rho})\cos(\varphi_I(\vec{\rho}) + \vec{g}\cdot\vec{\rho}) \, , \end{aligned} \tag{6}$$

where $\vec{g}$ is the in-plane component of the wave vector of the plane wave and is determined by the tilt angle introduced by the biprism. Three contributions to the intensity can be distinguished: the reference image intensity, the specimen image intensity and a set of cosinusoidal fringes, whose local phase shift and amplitude are given by the phase and amplitude of the electron wave function in the image plane. The phase and amplitude of the wave function of interest can be extracted from the hologram by applying a Fourier Transform (FT) and reconstructing the complex wave function by means of an inverse Fourier Transform (IFT). If required, $2\pi$ phase discontinuities can be removed. The FT of Eq. 3 can be written in the form

$$FT[I_{holo}(\vec{r})] = \delta(\vec{k}) + FT[A_I^2(\vec{r})] + \delta(\vec{k}+\vec{g}) \otimes FT[A_I(\vec{r})e^{i\varphi_I(\vec{r})}] + \\ + \delta(\vec{k}-\vec{g}) \otimes FT[A_I(\vec{r})e^{-i\varphi_I(\vec{r})}] \,, \qquad (7)$$

where $\delta(.)$ is the Dirac delta function and $f \otimes g$ represents the convolution of $f$ and $g$. In this expression, the first two terms are the FTs of the reference and sample wave function, respectively, located at $\vec{k}=0$. The last two terms are peaked at $\vec{k}=\pm\vec{g}$, correspond to the FTs of the desired image wave function and its complex conjugate and are known as *sidebands,* while those centered on the origin are referred to as a *center band*. The larger the value of $\vec{k}$, *i.e.*, the larger the tilt of the reference wave, the further from the origin are the sidebands. The sidebands contain both amplitude and phase information about the wave function of interest. In order to recover the complex wave function, one of the side bands is selected and isolated by applying a circular mask, shifted to the origin of reciprocal space and inverse Fourier transformed. The mask should have soft edges and a radius that is no larger than one third of the distance between the sideband and the origin (as the radius of the center band is twice that of the sideband) [2]. The phase image may need to be "unwrapped" to remove $2\pi$ phase discontinuities, which appear at positions where the phase shift exceeds $2\pi$, as IFT operations are calculated modulo $2\pi$ [3]. The number of phase wraps can sometimes be reduced by removing a constant phase gradient by repositioning the center of the sideband.

*1.2 Synthetic hologram generation*

Since the 1960s, as a result of advances in computational power, so-called *computer-generated holograms* (CGHs) have been introduced. As *Lesem et al.* [4] stated in 1968 when referring to holograms for 3-D displays:

*"A properly illuminated hologram forms for the viewer a picture which is identical with that which he would observe if he were looking at the scene himself. A computer generated hologram yields such a 3-D picture, without the original scene ever having to exist"*.

Based on this simple explanation, it is possible to understand how CGHs allow desired patterns to be designed and tested without the need to create models for each iteration, reducing the time required to make a synthetic CGH (S-CGH). The term "synthetic" is used to underline the fact that the last step involves producing a hologram that will be inserted into a microscope or an optical bench. Figure 1 shows a representation of the two concepts (or modes) of holography: "conventional" image holography (Fig. 1a) and "synthetic" holography (Fig. 1b), in this case for a plane wave incident on an S-CGH to obtain a desired wave function.

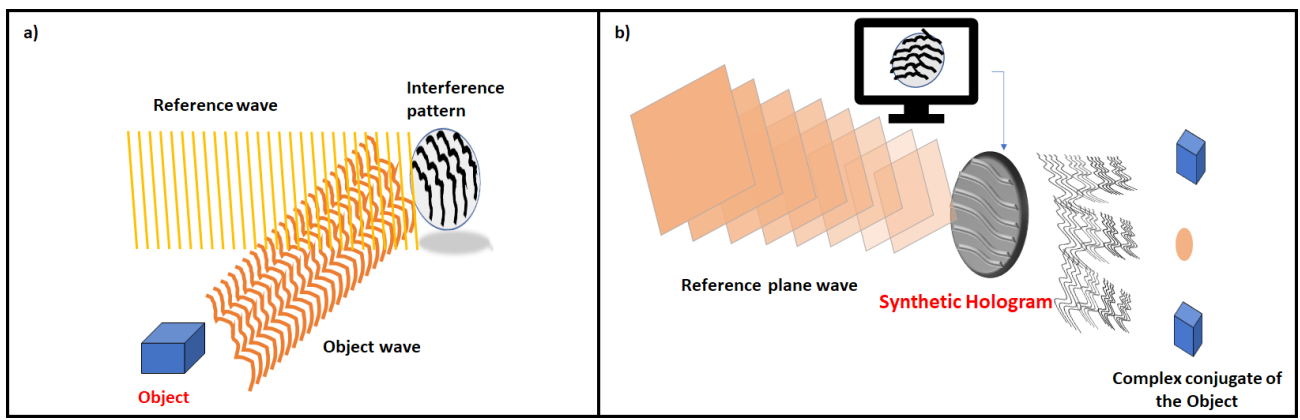

*Figure 1: Schematic diagram of a) traditional "image" holography and b) "synthetic" holography.*

Generation of the desired object and reference wave functions, as well as the interferometric process, can be carried out computationally. Most of the computational steps described in this article rely have been carried out using a modified version of *Stem_Cell* software [5]. In this software, the interferometric process, which consists of overlapping wave functions in a given plane, is carried out computationally. A CGH generated by such as set of operations (*i.e.*, the interference intensity pattern) is exported to a file, which can be used to fabricate an S-CGH.

The typical dimension of an S-CGH ranges from a few µm to hundreds of µm, while the smallest features can be only a few tens of nm in size. The fabrication process requires modern state-of-the-art machines and well-developed processes, which are described in the next sections. An S-CGH shows a desired function when it is illuminated by an incident (reference) beam. A manufactured S-CGH can be mounted in one of the condenser aperture planes of an electron microscope. However, as a periodicity of 100 nm corresponds to a scattering angle of only 1 µrad, the required very long focal length is usually not accessible when the main imaging lens (the objective lens) is switched on. The objective lens may therefore need to be switched off, for example by using the Low Mag mode of the TEM. When a synthetic hologram is positioned in the condenser system and used to form a focused probe, the resulting diffracted beams are visible in the specimen plane. In low-angle diffraction (LAD) mode, an S-CGH acts as a *transmission diffraction grating*, with each of the diffracted beams centered on a different position in the diffraction plane. A simple recipe for producing an S-CGH is to take the formula for an "image hologram" and to invert it. By illuminating a material that introduces the same amplitude (or intensity) modulation as in Eq. 6, it is then possible to obtain an object from which one of the diffracted beams corresponds to the wave that passed through the sample.

Depending on the type of interaction with the beam, synthetic holograms can be divided into three primary categories: (i) amplitude holograms; (ii) phase holograms; (iii) mixed (amplitude-phase) holograms. In this way, they can encode (i) only the phase, or (ii) both the phase and the amplitude of a wave function of interest. The character of the hologram (phase, amplitude or both) and the type of encoding are independent. For example, a phase hologram can encode both the amplitude and the phase of a wave function, but with some restriction on efficiency. With respect to an incident plane wave, the transmittance function of an (amplitude-phase) S-CGH can be written in the form

$$T_H(\vec{\rho}) = A(\vec{\rho})e^{i\Delta\varphi(\vec{\rho})} \ , \tag{8}$$

where $\vec{\rho}$ is the transverse spatial coordinate with respect to the propagation direction of the beam, while $A(\vec{\rho})$ and $\Delta\varphi(\vec{\rho})$ are amplitude and phase modulations. An amplitude hologram modifies the amplitude of the incident wave $A(\vec{\rho})$ and keeps the phase unchanged, such that $\Delta\varphi(\vec{\rho}) = constant$[6]; a phase hologram modifies the phase of the incident wave by modulating $\Delta\varphi(\vec{\rho})$, such that $A(\vec{\rho}) = constant$ [7]; a mixed hologram modifies both $A(\vec{\rho})$ and $\Delta\varphi(\vec{\rho})$ [8]. It should be noted that a "phase" S-CGH always has an additional absorption effect that depends on the thickness of the material and its chemical composition, while even a pure-phase S-CGH also has an effect on the amplitude of the wave function. An alternative way to achieve pure phase modification is to substitute a material-based hologram with a structured electric and/ or magnetic field, which introduces a desired phase modulation. It is then more challenging to design a complex and arbitrary phase shift [9]. This paper does not concentrate on such phase elements.

### 1.3 Different types of holograms

#### 1.3.1 Amplitude holograms

In light optics, binary holograms (characterized by a local transmittance that is 0 or 1) are produced from partially transparent elements, such as gratings that are made from metals or substrates that can block a light beam in some regions in the transverse plane. They are considered to be the simplest types of S-CGHs that can be fabricated. Amplitude modulation in transmission is usually achieved by covering parts of the beam with a material that can prevent light from passing through it (an opaque material that absorbs the beam), by deflecting the beam to a high angle or by reflecting part of the incident beam.

In electron optics, amplitude modulation mainly results from the combined effect of inelastic and high-angle scattering. Both scattering processes are usually stronger for heavy materials and thicker substrates. The blocking of electrons can be achieved by using a thick sputtered layer of a high-atomic-number element such as Au or Pt. By doing so, the wave front amplitude is locally fully preserved or completely blocked. Since a hologram absorbs or scatters electrons, its action is non-unitary and the overall intensity is reduced by a factor that is proportional to the blocked area in the incident beam cross-section. By definition, amplitude holograms block part of the electron beam and have limited efficiency. Since absorption modulation is an amplitude-dominated effect, such holograms result in diffraction pattern that is symmetrical between positive and negative orders. Moreover, it is impossible to concentrate the intensity on a single diffraction spot and a large part of the intensity is directed to the $0^{th}$ order transmitted beam. In order to gauge the absorption of a material, a useful parameter is the mean free path for plasmon excitation. The mean free paths of several materials are reported in Table 1 for 200 keV electrons.

| Material | Au | Ag | Pt | $Si_3N_4$ | $SiO_2$ | $Al_2O_3$ | a-C |
|---|---|---|---|---|---|---|---|
| Theoretical (nm) [10] | 76.1 | 88.3 | 76.4 | 135.3 | 133.6 | 135.7 | 106 |
| Experimental (nm) [11] | 84 (120) | 100 (125) | 82 (120) | | 155 | 140 | 160 |

Table 1: Theoretical and experimental mean free paths for 200 keV electrons. The experimental values are total inelastic mean free paths, which include single electron excitations such as inner-shell ionization edges. The terms in parentheses are mean free paths for collective valence electron (i.e., plasmon) excitations [10].

It should also be noted that the construction of pure amplitude holograms, in which absorptive material is alternated with vacuum, is complicated at small sizes because of the probability that long and thin parts of the hologram may collapse or join together during fabrication or under electron beam illumination.

**1.3.2 Phase holograms**

In light optics, one way to implement a phase modulation is by etching grooves of a desired structure on a (transparent or reflective) surface, in order for the optical path inside (or upon reflection from) the material to vary from one ray to another, thereby locally changing the phase of the outgoing wave function. In contrast, in electron optics phase modulation is achieved by exploiting the relationship between scalar and vector potentials. The phase shift of an electron wave function [12] is given by the expression

$$\Delta\varphi(\vec{\rho}) = C_E \int_{-\infty}^{+\infty} V(\vec{\rho},z)dz - \frac{e}{\hbar}\int_{-\infty}^{+\infty} A_z(\vec{\rho},z)dz \ , \tag{9}$$

where

$$C_E = \frac{2\pi}{\lambda}\frac{e}{E}\frac{E_0 + E}{2E_0 + E} \ , \tag{10}$$

$V(\vec{\rho},z)$ and $A_z(\vec{\rho},z)$ are the scalar electrostatic potential and the $z$ component of the magnetic vector potential, respectively, $e$ is the absolute value of the electron charge, $\hbar = h/2\pi$ is the reduced Planck constant, $\lambda$ is the relativistic electron wavelength, $E_0$ is the electron energy at rest and $E$ is the energy of the moving electrons. Typical electron energies in a TEM are 200 or 300 $keV$, resulting in corresponding values of $C_{E\_200keV} = 7.3 \times 10^{-3}\frac{rad}{V \cdot nm}$ and $C_{E\_300keV} = 6.6 \times 10^{-3}\frac{rad}{V \cdot nm}$. In a non-magnetic material, only the scalar electrostatic potential contributes to the phase shift. It can often be approximated by the mean inner potential $V_{mip}$, which provides a local acceleration to the electrons [13], modifying the electron-optical path. The phase variation due to $V_{mip}$ and the local thickness $t(\vec{\rho})$ is given by a simplified version of Eq. 9:

$$\Delta\varphi(\vec{\rho}) = C_E \int_0^{t(\vec{\rho})} V_{mip} dz = C_E V_{mip}\, t(\vec{\rho}) \ . \tag{11}$$

When choosing a material for a synthetic hologram, one must take into account the robustness, electrical conductivity and value of $V_{mip}$. Most phase S-CGHs are currently made using $Si_3N_4$, which can be used to fabricate a nearly pure phase mask since it is almost transparent to an incoming electron beam. S-CGHs can be obtained by "carving" grooves in a free-standing $Si_3N_4$ membrane. The calculated value of $V_{mip}$ for $Si_3N_4$ has been estimated to be $\approx 15\ V$ [14]. The same value was reported by Harvey *et al.* [15], a slightly higher value was reported by Bhattacharyya *et al.* [16], while a value of $\sim 10V$ was found by Shiloh *et al.* [17]. The specific SiN preparation process and tension may affect the precise value. On the assumption that $V_{mip} \approx 15\ V$, the required thickness to introduce a $2\pi$ phase shift for several electron energies is shown in Table 2.

| $E\ (keV)$ | 60   | 80   | 120  | 200  | 300  |
|------------|------|------|------|------|------|
| $t\ (nm)$  | 36.9 | 41.5 | 48.5 | 57.4 | 64.2 |

*Table 2: Si$_3$N$_4$ thickness required to introduce a 2π phase shift for different electron energies for $V_{mip} \approx 15\ V$.*

In recent years, new materials have been explored for the production of phase S-CGHs, with promising results shown for amorphous C [18]. The mean inner potentials of the materials from Table 1 are reported in Table 3.

| Material | Au | Ag | Pt | Si$_3$N$_4$ | SiO$_2$ | Al$_2$O$_3$ | a-C |
|---|---|---|---|---|---|---|---|
| Theoretical (V) | 25 ÷ 31 | 18.74 ÷ 23 | 20 ÷ 27 | 11.3 ÷ 17.6 | ~15.1 $V$ | 15.7 ÷ 16.7 [19] | 10.1 ÷ 11.3 [20] |
| Experimental (V) | 21 ÷ 30 | 17 ÷ 23 | ~25 $V$ | ~15 $V$ [14] | ~17 $V$ | 16.9 ± 0.36 [19] | 9.09 ÷ 10.7 [21] [22] |

*Table 3: Theoretical and experimental mean inner potentials of the materials in Table 1. Most values are taken from [23] and [24]. For amorphous C, the mean inner potential depends on the density of the material.*

A schematic diagram of the operating principle of a phase mask is shown in Fig. 2 for an incoming plane wave, whose wave front is shown in red. A phase ramp, which results from the thickness profile of the phase mask, introduces a linear phase shift to a plane wave in a specific direction or azimuthal angle. The thinner region does not alter the electron beam wave front significantly, while the thicker region can be chosen so that it introduces a phase shift such as 2π to an incoming electron beam. The phase-shifted wave front, which is shown using a colour gradient, causes the electron beam to be deflected (left) or to carry OAM (right). The phase masks shown in Fig. 2 are *in-line* S-CGHs and are the simplest types that can be designed, having no transverse grating. Nevertheless, great manufacturing precision is required, as the phase shift is encoded in the pointwise thickness profile of the material. Such an *in-line* S-CGH does not require a reference beam.

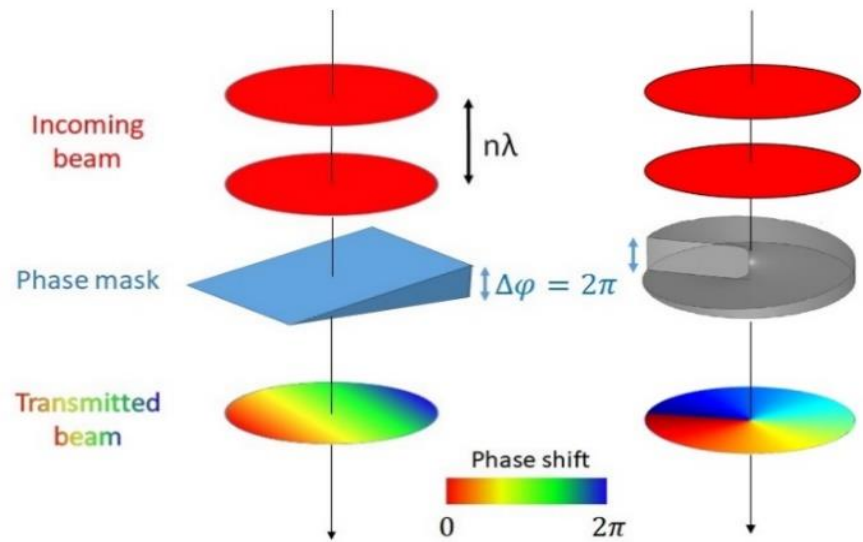

*Figure 2: Schematic diagram showing the phase shifting effect on an incident plane electron wave of (left) a phase ramp and (right) a spiral phase mask. Adapted from [10].*

*Off-axis* S-CGHs are obtained by using a plane wave as a reference, just as for an *off-axis* hologram. The primary advantage of using an *off-axis* S-CGH is that the encoded beam and hologram diffraction pattern are decoupled from one another. Off-axis S-CGHs are therefore less sensitive to small imperfections.

### 1.3.3 Amplitude-phase holograms

In the most general class of holograms, the material encodes a change in both the amplitude and the phase of the electron beam. In a strict sense, all phase holograms are amplitude-phase holograms, since any variation in thickness affects both the amplitude and the phase of the electron wave upon propagation [25]. Moreover, given the practical difficulties of fabricating complex amplitude gratings that are mechanically stable, amplitude gratings are often fabricated on continuous $Si_3N_4$ membranes [26]. Depending on how transparent the thick parts of the grating are, all levels between amplitude grating and phase grating can be obtained. The smart use of two materials can be used to cancel amplitude effects in a phase hologram, or to add an amplitude envelope to a phase grating. This approach could allow for joint amplitude and phase encoding of wave functions but has not been explored in detail. Even the aperture effect that encloses a phase hologram (and truncates the beam) is still a type of amplitude filtering. In general, amplitude and phase modulations have slightly different effects, and the two contributions are superimposed. It is therefore difficult to control the amplitude and phase simultaneously using different holograms.

### 1.4 Calculation of holograms

*1.4.1 Encoding the phase in phase holograms and amplitude holograms*

In this section, it is shown how to calculate a hologram for a target wave function for the off-axis case, *i.e.*, when a target wave function $\Psi(k)$ is reproduced in the first diffraction order. We assume that the reference wave is a plane wave with in-plane wave vector $\vec{g}$, as determined by the inclination of the reference wave.

In "imaging holography", the condition for good reconstruction is a large enough fringe spacing and an object with a narrow frequency band, in order to be able to isolate and demodulate the phase properly. Similarly, in synthetic holography, a desired function $\Psi(k)$ should have a compact support so that its extension in Fourier space is smaller than a reference frequency $\vec{g}$. The required phase modulation $\Delta\varphi(\vec{\rho})$ needs to be calculated based on the desired diffraction shape. The hypothesis is that one can control the phase $\alpha(\vec{\rho})$ of the desired diffracted beam at the plane of the hologram. This should simply be the phase of the inverse FT of the object $\Psi(k)$, with the addition of a phase gradient as a result of the off-axis tilt. One can numerically calculate $\alpha(\vec{\rho}) = arg\{FT^{-1}(\Psi(k))\} + \vec{g} \cdot \vec{\rho}$ . For example, $\alpha(\vec{\rho}) = \ell\theta + \vec{g} \cdot \vec{\rho}$ for a vortex beam, where $\ell$ is the desired winding number of the vortex and $\theta$ is the azimuth of the $\vec{\rho}$ coordinate in the hologram plane.

If $FT^{-1}(\Psi(k))$ had a constant amplitude, then $T_g = exp\left(i\alpha(\vec{\rho})\right)$ would be the transmission function $T_g$ of the desired phase plate and its Fourier transform would be $\Psi(k)$, apart from a tilt of $\vec{g}$. However, only the phase at the exit plane of the hologram is encoded. Methods to generalize this approach are discussed below. In the phase hologram case, the phase can be any function $\Delta\varphi = f(\alpha)$, with the periodicity condition $f(\alpha(\vec{\rho})) = f(\alpha(\vec{\rho}) + 2n\pi)$. For example, a sinusoidal grating that is used to generate vortex beams would be $\Delta\varphi = \varphi_0 sin\,(\ell\theta + \vec{g} \cdot \vec{\rho})$. Since the function $\alpha(\vec{\rho})$ is, by definition, bandwidth-limited, the transmission function of the full hologram is $T = exp\,(i\,f(\,\alpha(\vec{\rho})))$, with approximate periodicity $\vec{g}$. It has a diffraction pattern that is given by many well-separated beams centered at $n\vec{g}$ , where $n \in \mathbb{Z}$ is the diffraction order, and each diffracted beam can be spatially separated. For the first order beam, the hologram acts as the desired transmission function $T_g =exp\,(i\alpha(\vec{\rho}))$ . As $f$ changes, so does the distribution between diffraction orders. For any form of $f$, the first diffraction order is only affected by a phase effect $T_g =exp\,(i\alpha(\vec{\rho}))$.

For an amplitude hologram, it is possible to assume a simplified form of interference, where one retains only the cross term in Eq. 5. The simplest form of interference is just a cross term $T =cos\,(\alpha(\vec{\rho}))$, which is clearly an amplitude modulation. However, as in the case of a phase hologram, one can use any function of the form $T = f(\alpha(\vec{\rho}))$. Analogously to a phase hologram, a sinusoidal amplitude grating that generates a vortex takes the form $T = \frac{A_0}{2}(1 + sin(\ell\theta + \vec{g} \cdot \vec{\rho}))$, where positivity of the amplitude hologram is enforced. Even in this case, for any form of $f$ the first order beam is only affected by a phase effect $T_g = exp(i\alpha(\vec{\rho}))$. Therefore, amplitude and phase holograms are exactly the same at the level of individual diffracted beams. However, the phasing and amplitude ratio between diffraction orders is different. For example, the first diffraction order is in phase with the zero order for an amplitude hologram, whereas there is generally a dephasing close to $\pi/2$ for a phase hologram. Amplitude effects are discussed in sections 1.4.3. and 1.4.4.

It is also important to mention Fresnel holograms. In this case, the desired intensity is not obtained in the Fraunhofer plane, but in an intermediate (Fresnel) plane. Although the concept is identical, the Fourier transform is then substituted by the Fresnel integral and

$$\alpha(\vec{\rho}) = arg\{\Psi(k) \otimes P(-\Delta z))\} + \vec{g} \cdot \vec{\rho}, \tag{12}$$

where $\otimes$ is a convolution integral and $P(\Delta z) = \frac{1}{i\lambda\Delta z} exp\left(i\frac{\pi}{\lambda\Delta z}(x^2 + y^2)\right)$.

### 1.4.2 Diffraction efficiency and groove profile

A key parameter that defines the performance of an off-axis S-CGH is its diffraction efficiency, which can be defined as the ratio between the intensity measured in a specific diffraction order and the intensity of the incoming beam or the total transmitted intensity. According to the first definition, the efficiency is

$$\eta_n^{(i)} = \frac{I_n}{I_{inc}}, \tag{13}$$

where $I_n$ is the intensity of order $n$ and $I_{inc}$ is the intensity of the incident beam. $\eta_n^{(i)}$ is then known as the absolute diffraction efficiency. The total transmitted efficiency is

$$I_{trans} = \sum_n I_n \tag{14}$$

and the second definition of diffraction efficiency is

$$\eta_n^{(t)} = \frac{I_n}{I_{trans}}. \tag{15}$$

In amplitude and phase S-CGHs, the beam has to pass through a material and the total transmitted intensity is reduced, typically by up to 50% and 30-40%, respectively. The reduction results from absorption and other inelastic processes, even in high-transmittance materials such as Si$_3$N$_4$. It has been suggested to use high-brightness electron sources to mitigate this problem. Henceforth, we use $\eta$ to refer to $\eta_n^{(t)}$, the so-called transmitted efficiency. When a distinction is necessary, the appropriate symbol is used. As mentioned above, the efficiency of an S-CGH depends on whether it is phase-modulated or amplitude-modulated. However, the efficiency also depends on the groove profile/ thickness pattern of the hologram. In order to establish the relationship between groove profile and efficiency, we begin by explaining how an incoming wave function is transformed after its interaction with an S-CGH. This interaction depends on the groove pattern. According to Eq. 8, the transfer function $T_H(\vec{\rho})$ describes the amplitude and phase of a beam that has passed through a diffraction grating. An alternative representation of the transfer function, specifically for a phase S-CGH, is given by the expression

$$T_H(\vec{\rho}) = e^{i\widetilde{V}t(\vec{\rho})} , \qquad (16)$$

where $\widetilde{V} = C_E V_{mip} + i\gamma$ is the complex index of refraction, $\gamma = \frac{1}{\lambda_{mfp}}$ is the absorption coefficient, $\lambda_{mfp}$ is the mean free path of an electron and $t(\vec{\rho})$ is the thickness profile. Since the transfer function is independent of the incident wave function $\Psi_{inc}(\vec{\rho})$, the transmitted wave function can be written in the form

$$\Psi_t(\vec{\rho}) = \Psi_{inc}(\vec{\rho})T(\vec{\rho}) = \Psi_{inc}(\vec{\rho})e^{i\widetilde{V}t(\vec{\rho})} . \qquad (17)$$

In most case studies, the incoming wave is assumed to be a plane wave and can be ignored, as it has a flat-phase wave front and its FT depends mainly on the transfer function. A generic diffraction grating is characterized by a periodic wave function $f(\alpha)$, which describes its groove pattern. It is usually dimensionless and normalized from zero to unity with period $2\pi$, such that $f(\alpha + 2\pi) = f(\alpha)$. The function can be expanded as a Fourier series in the form

$$f(\alpha) = \sum_{n=-\infty}^{\infty} c_n e^{in\alpha} , \qquad (18)$$

where $n \in \mathbb{Z}$ and the $n^{\text{th}}$ Fourier coefficient

$$c_n = \frac{1}{2\pi} \int_0^{2\pi} f(\alpha) e^{-in\alpha} d\alpha . \qquad (19)$$

Each value of $n$ represents one diffraction order. If $f(\alpha)$ is real-valued, then $c_n = c^*_{-n}$, where the asterisk denotes a complex conjugate and $c_0$ is real. The Fourier power spectrum of $f(\alpha)$ is given by the expression

$$S = \sum_n |c_n|^2 . \qquad (20)$$

For a bi-dimensional grating with a single type of groove profile, the periodic function $f(\alpha)$ has

$$\alpha = \alpha(\vec{\rho}) , \qquad (21)$$

where $|\vec{\rho}|$ and $\theta$, the azimuthal angle, are polar coordinates that define the grating and $|\vec{\rho}|$ is measured in units of the grating spatial period $\Lambda$ in the $\theta = 0$ direction. For most functions $\alpha(\vec{\rho})$ and $f(\alpha(\vec{\rho}))$ that are used for diffraction gratings, the Jacobian is the same for each diffraction order (except for the zeroth order) and can be disregarded in the calculations. A detailed mathematical analysis is not included in this tutorial.

For an amplitude S-CGH, the field transmission Fresnel function $T(x, y)$ is proportional to the grating function

$$T(\vec{\rho}) = b\, f(\alpha(\vec{\rho})) , \qquad (22)$$

where $b$ is a constant and $0 \leq b \leq 1$.

The wave function impinging on the S-CGH is denoted $\Psi_{in}(\vec{\rho})$. The output wave function can then be determined as follows:

$$\Psi_{out}(\vec{\rho}) = T(\vec{\rho})\Psi_{in}(\vec{\rho}) = b\, f(\alpha(\vec{\rho}))\Psi_{in}(\vec{\rho}) . \qquad (23)$$

For a phase S-CGH, the transmission function is given by the expression

$$T(\vec{\rho}) = e^{i\tilde{a}f(\alpha(\vec{\rho}))} , \qquad (24)$$

where $\tilde{a} = a_1 + ia_2$ is a complex number, $a_1 = C_E V_{mip} t_M$, $a_2 = \gamma t_M$ and $t_M$ is the maximum thickness difference between a peak and a valley. It can be shown that, depending on the microscope accelerating voltage, $a_2 \sim 7 \div 8\%\, a_1$ for $Si_3N_4$. Furthermore, the product of $t_M$ and $f(\alpha(\vec{\rho}))$ yields the local thickness of the grating $t(\vec{\rho})$. Therefore, for this type of hologram the output wave function is given by the expression

$$\Psi_{out}(\vec{\rho}) = \Psi_{in}(\vec{\rho}) e^{i\tilde{a} f(\alpha(\vec{\rho}))} \ . \qquad (25)$$

The efficiency of the S-CGH can be estimated/ calculated from the power transmission spectrum, given by the sum of the Fourier coefficients of the transmission function Fourier series expansion

$$\mathcal{T}(\vec{\rho}) = \sum_n |\tau_n|^2 \ , \qquad (26)$$

where the intensity of the $n$-th diffraction order is modulated by the transmission coefficient $|\tau_n|^2$. For amplitude S-CGHs, $\theta_n = \frac{n\lambda}{\Lambda}$ is the diffraction angle for the $n$-th order diffracted beam. The power transmittance is unity for a pure phase grating, so it can ideally support a very high power peak.

### *1.4.3 Comparison between grating profiles*

In this section, a series of grating profiles is presented for both amplitude and phase S-CGHs. In each case, the grating profile function and the Fourier coefficients of the transmission function $\tau_n$ are given. Calculations describing how the equations were obtained are given in Appendix A.

**Sinusoidal/ cosinusoidal profile**

The simplest profile is sinusoidal/ cosinusoidal. As they have the same characteristics, only one is considered. For an amplitude S-CGH with a cosinusoidal profile, $f(\alpha) = \frac{1}{2}(1 + cos(\alpha(\vec{\rho})))$. The transmission function is

$$T(\vec{\rho}) = \frac{b}{2}(1 + cos(\alpha(\vec{\rho}))) \ , \qquad (27)$$

where $0 \leq b \leq 1$ is a constant. For a phase S-CGH, the transmission function can be written in the form

$$T(\vec{\rho}) = e^{i\frac{\tilde{a}}{2}(1+\cos(\alpha(\vec{\rho})))} =$$
$$= e^{i\frac{a_1}{2}\cos(\alpha(\vec{\rho}))} e^{-\frac{a_2}{2}\cos(\alpha(\vec{\rho}))} e^{i\frac{\tilde{a}}{2}}$$
$$= e^{ia'_1 \cos(\alpha(\vec{\rho}))} e^{-a'_2 \cos(\alpha(\vec{\rho}))} e^{i\tilde{a}'} \quad . \tag{28}$$

For both types of holograms, $f(\alpha)$ is normalized between zero and unity, so that for an amplitude S-CGH the power transmittance changes locally between 0 for full absorption and 1 for no absorption. For an ideal phase S-CGH, the power transmittance is always 1 and it is possible to estimate the optimal phase shift introduced by the local thickness profile to maximize the intensity of one of the diffraction orders (usually $n = \pm 1$). For an amplitude S-CGH, the squared modulus of the Fourier coefficient of the transmission function is

$$|\tau_n(\vec{\rho})|^2 = \begin{cases} \frac{1}{4} b^2 \text{ for } n = 0 \\ \frac{1}{16} b^2 \text{ for } n = \pm 1 \\ 0 \text{ for the other orders} \end{cases}, \tag{29}$$

for the $n^{th}$ order, whereas for a phase S-CGH the corresponding expression is

$$|\tau_n(\vec{\rho})|^2 = J_n^2(\tilde{a}') e^{-2a_2'}$$
$$\approx \left[ J_n^2(a'_1) - \frac{{a'_2}^2}{4} \left( J_{n-1}(a'_1) - J_{n+1}(a'_1) \right)^2 \right] e^{-2a'_2} \quad . \tag{30}$$

Figure 3 shows the characteristic efficiency of the first diffracted order for a phase S-CGH plotted as a function of the parameters $a'_1$ and $a'_2$. The efficiency reaches a maximum when $a'_1 \approx 1.84 \, rad$. In the "real phase S-CGH" profile, the contribution of absorption is appreciable, with the main contribution originating from the exponential term $e^{-2a_2'}$. The ideal value of $a'_1$, which maximizes $|\tau_1(\vec{\rho})|^2$ of an ideal phase S-CGH, corresponds to a peak-to-valley phase difference of ~3.68 radians (1.17$\pi$), which corresponds to $t_M \approx 33.6$ nm for 200 keV electrons.

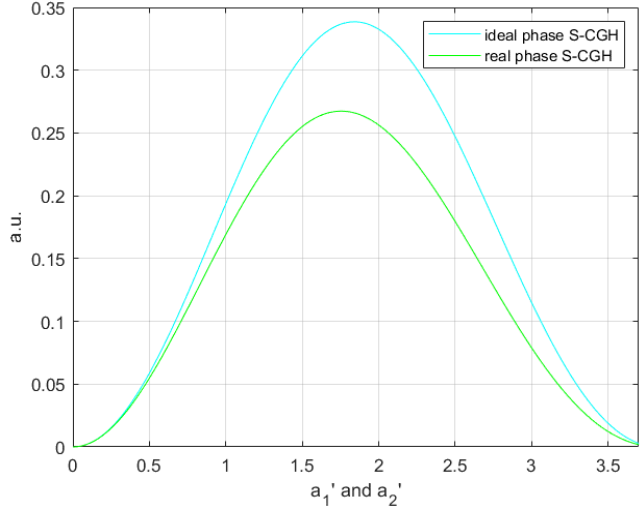

*Figure 3: $|\tau_1|^2$ plotted for a phase S-CGH with a cosinusoidal profile (Fig. 4) as a function of $a'_1$.*

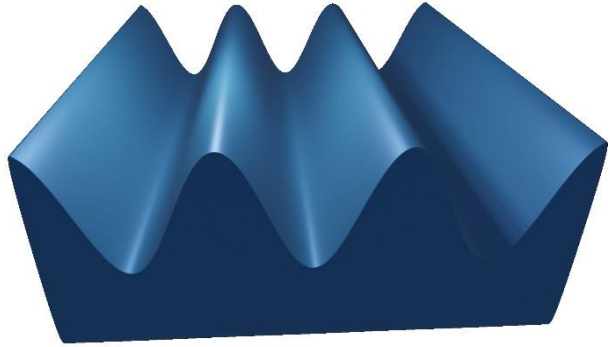

*Figure 4: Three-dimensional rendering of a cosinusoidal profile.*

**Squared profile**

The second profile considered here is a squared profile, for which $f(\alpha) = \frac{1}{2}(1 + Sign(sin(\alpha(\vec{\rho}))))$. For an amplitude S-CGH, the transmission function takes the form

$$T(\vec{\rho}) = \frac{b}{2}\Big(1 + Sign(sin(\alpha(\vec{\rho})))\Big) , \qquad (31)$$

while for a phase S-CGH the transmission function can be written

$$T(\vec{\rho}) = e^{i\frac{\tilde{a}}{2}(1+Sign(sin(\alpha(\vec{\rho}))))} =$$
$$= e^{i\tilde{a}'Sign(sin(\alpha(\vec{\rho})))} e^{i\tilde{a}'} =$$
$$= e^{ia'_1 Sign(sin(\alpha(\vec{\rho})))} \cdot e^{-a'_2 Sign(sin(\alpha(\vec{\rho})))} e^{i\tilde{a}'} . \qquad (32)$$

For an amplitude grating, the square modulus of the Fourier coefficients of the transmission function is

$$|\tau_n(\vec{\rho})|^2 = \begin{cases} \frac{1}{4}b^2 & \text{for } n = 0 \\ \frac{1}{n^2\pi^2}b^2 & \text{for } n = \text{odd} \\ 0 & \text{for } n = \text{even} \end{cases} \quad , \tag{33}$$

while for a phase grating it is

$$|\tau_n(\vec{\rho})|^2 = \begin{cases} [\cos^2(a'_1)\cosh^2(a'_2) + \sin^2(a'_1)\sinh^2(a'_2)]e^{-2a_2'} & \text{for } n = 0 \\ \frac{4}{n^2\pi^2}[\sin^2(a'_1)\cosh^2(a'_2) + \cos^2(a'_1)\sinh^2(a'_2)]e^{-2a_2'} & \text{for } n = \text{odd} \\ 0 & \text{for } n = \text{even} \end{cases} \quad . \tag{34}$$

Figure 5 shows the efficiency of the first diffraction order for a phase S-CGH. The maximum is reached when $a'_1 \approx 1.57\ rad$, so the optimal peak-to-valley phase difference corresponds to $\Delta\varphi \approx \pi$ for an ideal phase S-CGH.

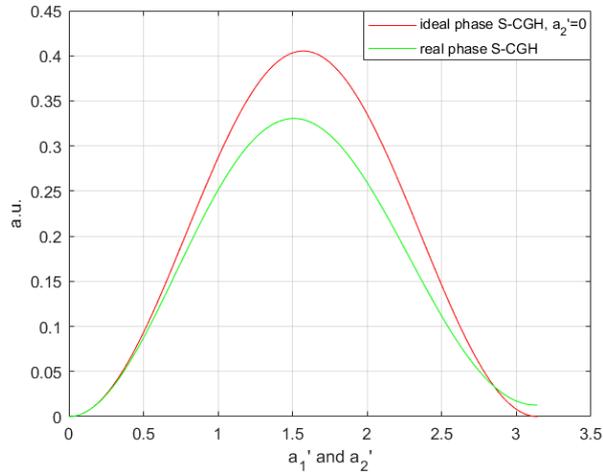

Figure 5: $|\tau_1|^2$ for a squared profile (Fig. 6) as a function of the amplitude of the sign function $a'_1$, corresponding to half of the peak-to-valley distance in radians. Red: no absorption. Green: with absorption.

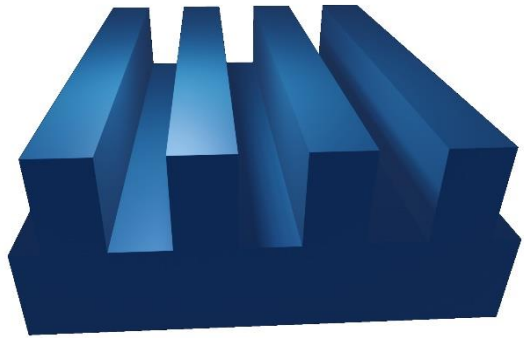

Figure 6: Three-dimensional rendering of a squared profile

**Triangular profile**

The third case is a triangular profile, which can be described (for an isosceles triangle) by the function $f(\alpha) = \frac{1}{\pi}(Sign(sin(\alpha(\vec{\rho}))))((\pi - Mod(\alpha(\vec{\rho}), 2\pi))$. For an amplitude S-CGH, the transmission function is

$$T(\vec{\rho}) = b\frac{1}{\pi}\left(Sign(sin(\alpha(\vec{\rho})))\right)\left(\pi - Mod(\alpha(\vec{\rho}), 2\pi)\right) , \qquad (35)$$

while for a phase S-CGH it is

$$T(\vec{\rho}) = e^{i\tilde{a}\frac{1}{\pi}\left(Sign(sin(\alpha(\vec{\rho})))\right)(\pi - Mod(\alpha(\vec{\rho}), 2\pi))} , \qquad (36)$$

where $Mod(p, q)$ is the remainder after dividing $p$ by $q$. For an amplitude S-CGH with a triangular modulation, the efficiency of the $n^{th}$ diffracted order is proportional to

$$|\tau_n|^2 = \begin{cases} \frac{1}{4}b^2 & for\ n = 0 \\ \frac{4}{n^4\pi^4}b^2 & for\ n = odd \\ 0 & for\ n = even \end{cases} , \qquad (37)$$

while for a phase S-CGH with a triangular modulation it is

$$|\tau_n|^2 = \\ = \frac{(a_1^2 + a_2^2)[1 + 2(-1)^{n+1} e^{-a_2}(\cos(a_1)) + e^{-2a_2}]}{[a_1^4 + 2a_1^2 a_2^2 + a_2^4 + (n\pi)^4 - 2n^2 a_1^2 \pi^2 + 2n^2 a_2^2 \pi^2]} . \qquad (38)$$

Figure 7 shows how the efficiency changes as different parameters are varied. If $a_2$ is non-zero, *i.e.*, if absorption is considered, then the efficiency is reduced and the peak moves to lower values of $a_1$. $|\tau_1|^2$ has a maximum at $a_1 \approx 4.31\ rad$ for an ideal phase S-CGH, whereas it is at $a_1 \approx 4\ rad$ for a real S-CGH.

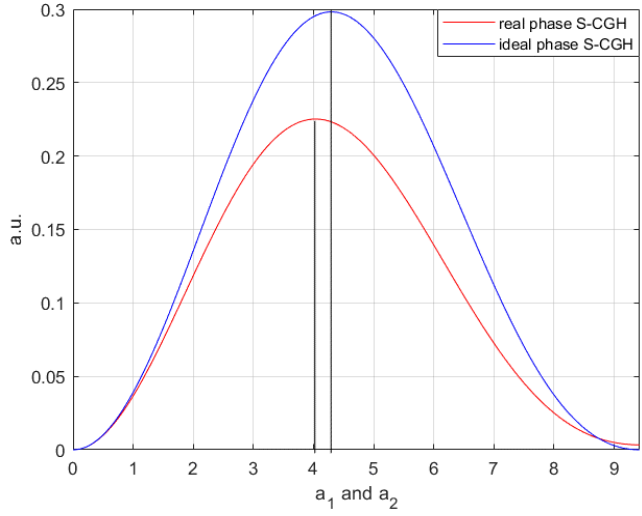

*Figure 7: $|\tau_1|^2$ for a triangular profile (Fig. 8) plotted for a phase S-CGH as a function of $a_1$ and $a_2$.*

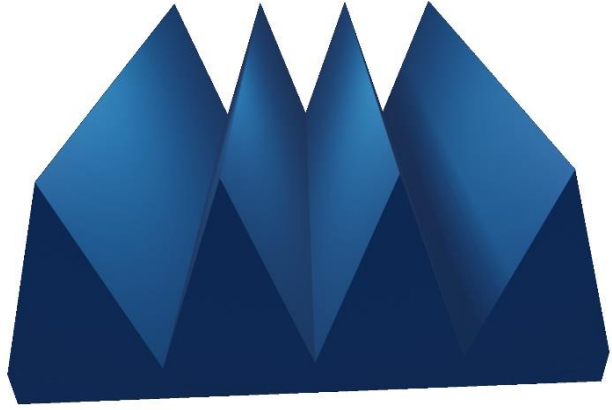

*Figure 8: Three-dimensional rendering of a triangular profile*

**Blazed profile**

An interesting triangular profile is a blazed profile, which is similar to a sawtooth blade and can be described by the function $f(\alpha) = \frac{1}{2\pi}\big(Mod(\alpha(\vec{\rho}), 2\pi)\big)$. For an amplitude S-CGH, the transmittance function is

$$T(\vec{\rho}) = b\frac{1}{2\pi}\big(Mod(\alpha(\vec{\rho}), 2\pi)\big) \,, \tag{39}$$

while for a phase S-CGH it is

$$T(\vec{\rho}) = e^{i\tilde{a}\frac{1}{2\pi}(Mod(\alpha(\vec{\rho}),2\pi))} \,. \tag{40}$$

For an amplitude S-CGH, the efficiency of the $n$-th order of diffraction is

$$|\tau_n|^2 = \begin{cases} \dfrac{1}{4} \ for\ n = 0 \\ \dfrac{1}{4\pi^2 n^2}\ for\ n \neq 0 \end{cases}, \tag{41}$$

while for a phase S-CGH it is

$$|\tau_n|^2 = \frac{(1 + e^{-2a_2} - 2\cos(a_1)e^{-a_2})}{[(a_1 + 2\pi n)^2 + a_2^2]}. \tag{42}$$

Figure 9 shows two features of a blazed profile for a phase S-CGH. First, in the ideal case maximum efficiency is reached when the peak-to-valley distance is equivalent to a phase difference of $2\pi$. Second, by tuning the shape it is possible to reach an efficiency of almost 100% in one of the first diffraction orders (Fig. 10). This profile is the only one that theoretically allows for the whole transmitted wave to be directed to one of the first diffraction orders.

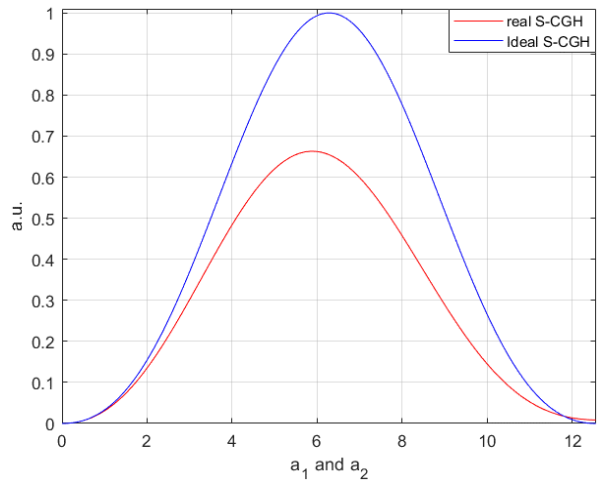

*Figure 9: Relative transmitted efficiency in the -1 diffracted order for a blazed profile with and without absorption. The maximum relative efficiency is reached for $a_1 \sim 2\pi$, meaning that when no absorption is considered ideal efficiency is obtained when the peak-to-valley phase difference due to thickness is $2\pi$.*

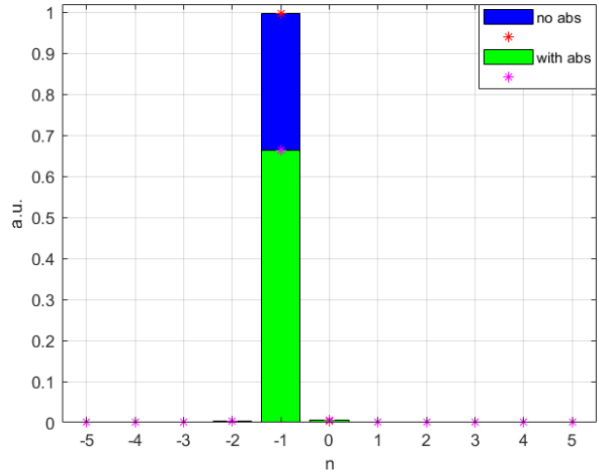

*Figure 10: Comparison between $|\tau_n|^2$ for different diffraction orders for the ideal case of a blazed phase S-CGH and a "real case" with absorption. In the ideal case, only the minus one order survives and all others ideally have zero intensity. If absorption is considered, $\mathcal{T}$ is no longer unitary.*

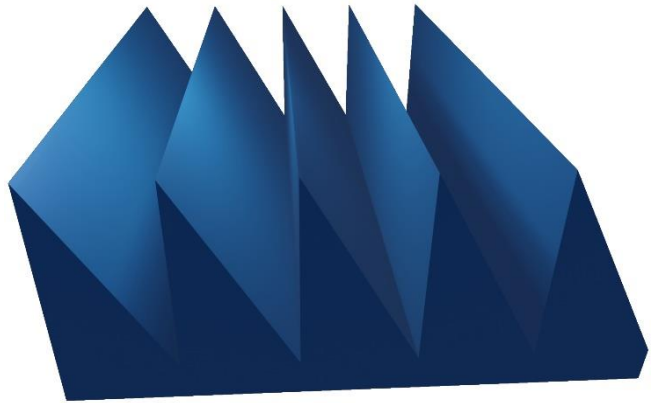

*Figure 11: Three-dimensional rendering of a blazed profile.*

If a blazed profile is not perfect and the grooves are more similar to scalene triangles, then the transmitted intensity is no longer concentrated in one of the first diffracted orders, but spreads to others. This is usually visible when looking at a diffraction pattern of a real blazed hologram affected by fabrication limitations. A more in-depth analysis regarding the optimization of a real blazed phase S-CGH is described in Section 3.1.3

### *1.4.4 Efficiencies of the profiles*

The efficiencies of the profiles that have been described are now compared, distinguishing between amplitude and phase S-CGHs. Efficiency is one of the critical parameters to consider during the design of a synthetic hologram. Here, one histogram is shown for each groove pattern, with the diffraction order on the horizontal axis and the transmitted efficiency $\eta_n^{(t)}$ on the vertical axis. For ease of visualization, only orders between -5 and +5 are shown.

**Amplitude S-CGH**

Figure 12 shows the intensity distribution between diffraction orders for different profile shapes (sinusoidal, squared, triangular and blazed) for an amplitude S-CGH. The central or zeroth order peak always has the highest efficiency. The total transmitted intensity is never 100%, since the hologram absorbs some incoming electrons. To a first approximation, if only absorption from the opaque part is considered and that from the supporting layer is neglected, the best performing shape is the squared profile, for which 50% of the intensity is transmitted. The worst performing shape is the blazed profile, for which only 33% is transmitted.

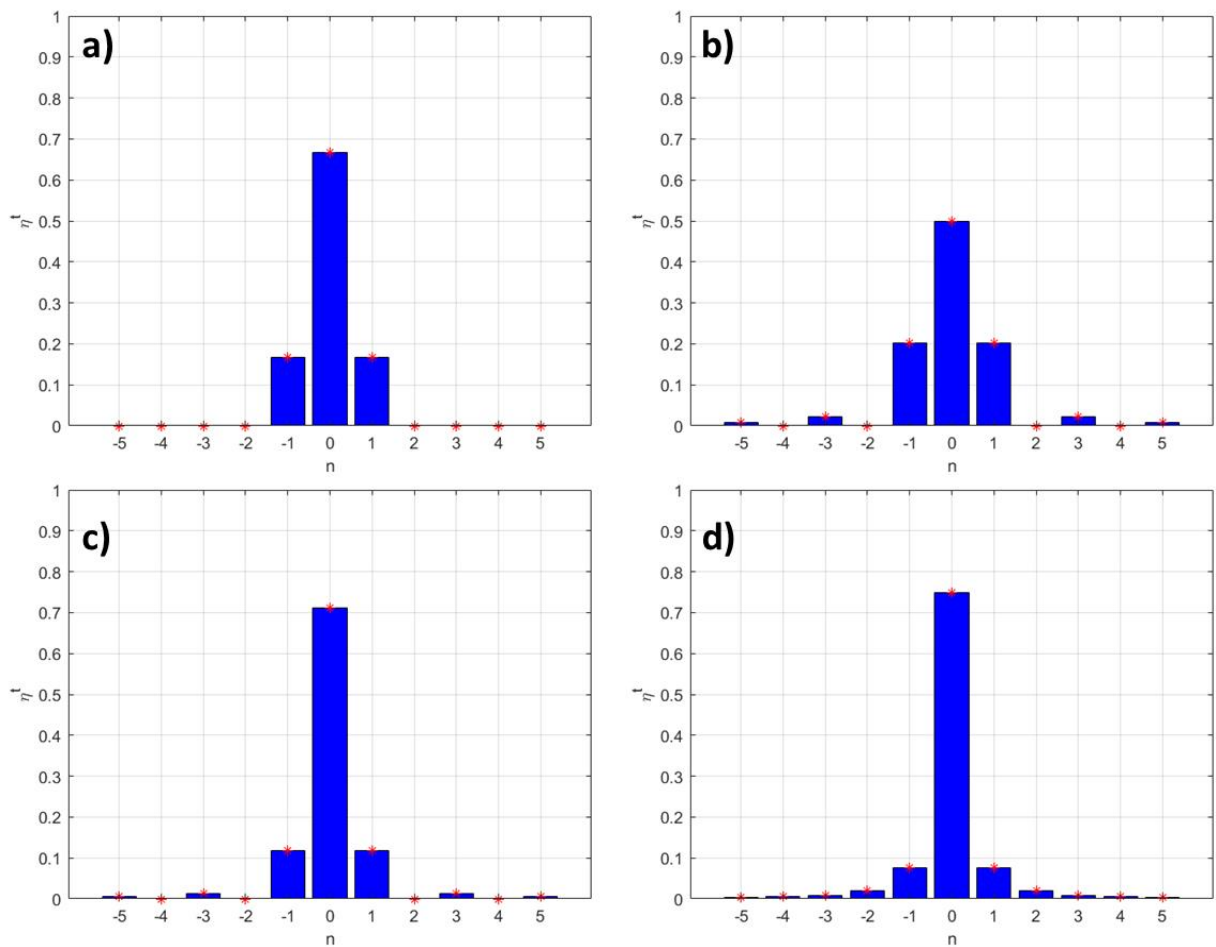

Figure 12: Transmitted efficiency of diffraction orders $n \in [-5,5]$ for an amplitude S-CGH:
a) sinusoidal profile; b) squared profile; c) triangular profile; d) blazed profile.

**Ideal phase S-CGH**

Figure 13 shows the intensity distribution between diffraction orders for different profile shapes for a phase S-CGH. The calculations have been carried out such that the phase difference maximizes the intensity in one of the first two diffraction orders. The zeroth order peak is always less intense than the first orders. While the phase difference between peak and valley can be tuned in a phase S-CGH, this is not possible for an amplitude S-CGH, for which the zeroth diffraction order is always the most intense. For an ideal phase S-CGH, in which absorption is omitted, the total transmitted intensity is almost 100% for all of the shapes considered here. Key values of efficiencies are reported in Table 4.

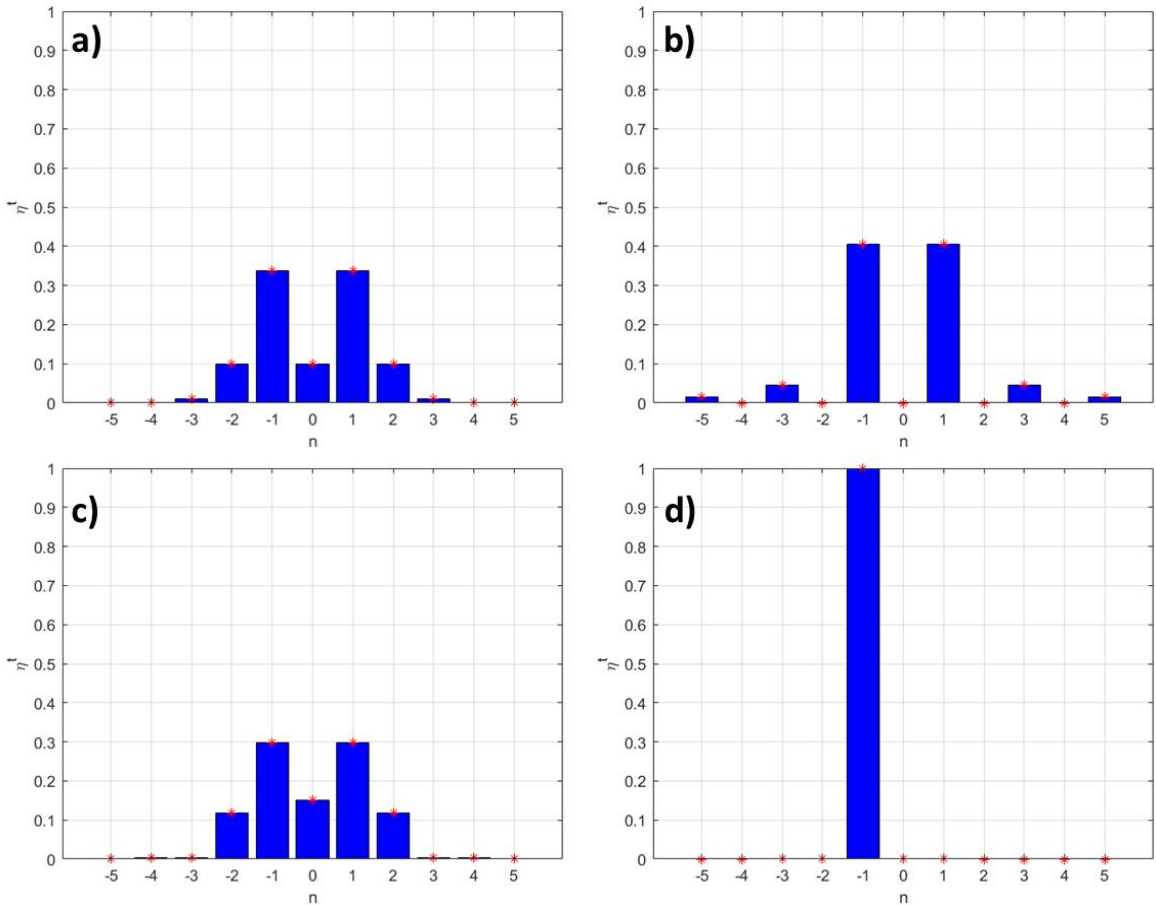

Figure 13: Transmitted efficiency of diffraction orders $n \in [-5,5]$ for an ideal phase S-CGH:
a) sinusoidal profile; b) squared profile; c) triangular profile; d) blazed profile.

| Profile shape | $\mathcal{T}_{amp}$ | $|\tau_{\pm 1}|^2$ | $\dfrac{|\tau_{\pm 1}|^2}{\mathcal{T}_{amp}}$ | $\mathcal{T}_{phase}$ | $|\tau_{\pm 1}|^2$ | $\dfrac{|\tau_{\pm 1}|^2}{\mathcal{T}_{phase}}$ |
|---|---|---|---|---|---|---|
| **Cosinusoidal** | 37.5% | 6.25% | 16.67% | 100% | 33.86% | 33.86% |
| **Squared** | 50% | 10.13% | 20.26% | 100% | 40.53% | 40.53% |
| **Triangular** | 35.13% | 4.11% | 11.69% | 100% | 29.82% | 29.82% |
| **Blazed** | 33.3% | 2.53% | 7.60% | 100% | 100% | 100% |

Table 4: Transmission power function, $n = \pm 1$ square modulus of Fourier coefficients and transmitted efficiency for two kinds of S-CGH and all of the groove profiles considered here.

### 1.5 Encoding both amplitude and phase in a synthetic hologram

### 1.5.1 Encoding amplitude and phase in a phase hologram

Unlike the other S-CGHs presented so far, which were aimed at generating a desired wave function in all non-zero diffraction orders (apart from a multiplicative factor for the angular momentum), mixed holograms generate a desired wave function in only one specific diffraction order. The method is based on tuning the peak-to-valley phase difference in each region of the S-CGH profile. This yields a local change in efficiency, which changes the wave front phase at the exit of the hologram, resulting in a change in the intensity of the beam. If $A(\vec{\rho})$ and $\varphi(\vec{\rho})$ are the amplitude and phase of a desired wave function, $B(\vec{\rho})$ is a normalized bounded positive function of amplitude, $C(\vec{\rho})$ is an analytical function of the amplitude and phase profiles of the desired field and $\Lambda$ is the period of the diffraction grating, then the profile to be fabricated takes the form [8]

$$T_{Mix}(\vec{\rho}) = exp\left[iB(\vec{\rho})Mod\left(C(\vec{\rho}) + \frac{2\pi\rho_{\theta=0}}{\Lambda}, 2\pi\right)\right],  \quad (43)$$

where

$$B(\vec{\rho}) = 1 + \pi^{-1}sinc^{-1}(A(\vec{\rho})),  \quad (44)$$

$$C(\vec{\rho}) = \varphi(\vec{\rho}) - \pi B(\vec{\rho}).  \quad (45)$$

and $sinc^{-1}()$ is the inverse of sinc function in the interval of $[-\pi, 0]$. Areas characterized by a full $2\pi$ phase shift contributes to the amplitude of the +1$^{st}$ diffraction order, whereas other areas spread intensity over other orders, limiting the intensity of the 1$^{st}$ order. The beam of interest is generated with the correct phase and amplitude information only in the 1$^{st}$ diffraction order.

### 1.5.2 Encoding amplitude and phase in an amplitude hologram

The approach described in section 1.5.1 for phase-only holograms is based on modulation of the peak-valley value. This modulation locally varies the efficiency of the grating and, therefore, the amplitude encoding. The same method can be used to achieve amplitude and phase encoding using an amplitude hologram.

For the sake of simplicity, we start by considering a binary mask (*i.e.*, a rectangular profile) and first encode the phase, before adding a modulation to the width of the groove that is related to the local efficiency of the hologram, as outlined in Eq. 32.

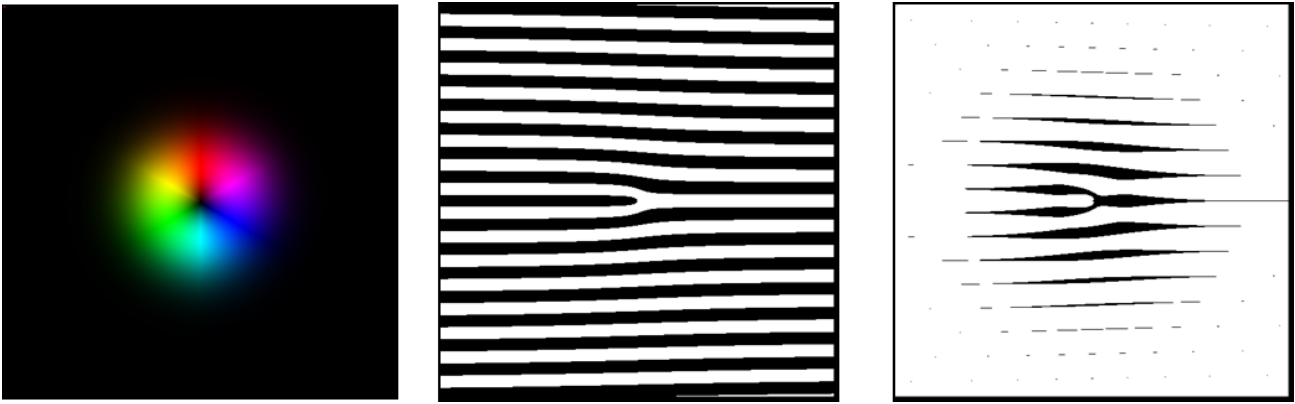

*Figure 14: Example of encoding the amplitude and phase of a Laguerre Gauss beam (ℓ = 1, p = 0) in an amplitude hologram. The phase encoding gives the pitchfork, but the change in groove width gives the amplitude envelope.*

In simple terms, the center of each groove is related to the phase modulation, while the width is related to the amplitude of the wave. Instead of a rectangular groove, one can choose any profile.

If a phase-only modulation is chosen such that $f(\alpha) \propto cos\,(kx + \alpha(\vec{\rho}))$, then the center of the fringes corresponds to the condition $cos\,(kx + \alpha(\vec{\rho})) = 1$. An amplitude modulation can be achieved by substituting the 1 with a "bias" function of the form $cos\,(q(\vec{\rho}))$, where $q(r)$ is a function of the local desired efficiency. The relation $cos(kx + \alpha(\vec{\rho})) = cos(q(\vec{\rho}))$ can then be used to find the clipping points at the sides of the groove. Further mathematical approaches based on this principle are possible (*e.g.*, [27]).

This conceptual scheme can be extended to include ideas for phase hologram encoding of amplitude and phase, as seen above. Furthermore, the above approach is more exact, as it accounts for the amplitude modulation effect on the phase shift and the phase effect on the amplitude.

*1.6 Sampling effect and choice of groove shape*

When deciding on the design of a hologram and particularly on the groove shape, the practical problem of the limited number of addressable or calculated pixels should be considered. Typically, a hologram uses a square with between 1000 and 4000 pixels on each side. Beyond 8000 pixels, it is computationally and experimentally demanding to build an S-CGH. A typical groove is sampled with $n_{pg} = 5$ to 20 pixels. If the resolution in the groove positioning is given approximately by $1/n_{pg}$, then the phase is defined to be within $2\pi/n_{pg}$. The resulting problem in phase shaping is greatest for the rectangular groove and, to some extent, for the blazed groove, as any discontinuity is defined by the size of the pixel. In contrast, a sinusoidal groove has the advantage that each pixel intensity defines the phase with no discontinuity. In other words, even if the center of a groove is not defined by a single pixel it can be calculated with sub-pixel precision as a weighted position average, whereas for a rectangular groove the phase is defined only on a discrete grid. A falsely encoded phase profile can result in additional intensity between the diffraction orders in the hologram's Fourier transform. Under specific conditions, it is possible to recognize many $(n_{pg})$ copies of the same beam. This effect is related to the Talbot effect [28]. A second point is the bandwidth of the function to be encoded. The carrier frequency $|\vec{g}|$ must be larger than the bandwidth of the signal. For a sinusoidal pattern, at least 4 pixels are needed per period. If the bandwidth is $B$, then $|\vec{g}| \gg 2B$ and

$$n_p = K_{max} \gg 8B \ . \tag{46}$$

For the case of a vortex beam with a top hat amplitude cutoff, $B \approx a\,\ell$ with $a \approx 1/\pi$, so for a beam with $\ell = 1000$ approximately 4000 pixels are required. A different groove shape could result in a different maximum winding number for the vortex that can be generated. One should also consider the fact that a groove shape depends on the fabrication approach. When using EBL, it is more difficult to fabricate a groove that is not rectangular. Further details about vortex beam generation and fabrication techniques are given below.

**Chapter 2 - Production of holograms: Electron beam lithography and focused ion beam milling**

The final step of S-CGH production is fabrication of the designed pattern on a chosen substrate. The most common substrate of choice is currently silicon nitride ($Si_3N_4$), while the two fabrication techniques that are typically used to make S-CGHs are focused ion beam (FIB) milling and electron beam lithography (EBL).

*2.1 Focused ion beam milling*

FIB milling is a powerful tool for the fabrication of designed patterns. A FIB instrument is used to generate a focused high-energy beam of accelerated ions, which are then directed towards a sample surface to remove material by sputtering. Although Ga ions are the most widely used ions for this purpose, Au, Ir, Ar, He, Xe, O, N and Si ions are also available. A higher-atomic-number element provides a higher milling yield, whereas a lower-atomic-number element offers greater accuracy in reproducing a desired pattern. FIB milling exploits so-called *knock-on sputtering*. For this to happen, the ion needs to be accelerated by a potential in the 1-50 kV range [29]. During FIB milling, an incoming ion hits a surface atom and transfers part of its kinetic energy to it, such that the atom is displaced from its equilibrium lattice position and collides with neighboring atoms, which can result in their release from the substrate. The incoming ion after several impacts loses almost entirely its primary energy and can be trapped in the target substrate, leading to ion implantation and a change in the properties of the target substrate. Although FIB ions can themselves also be exploited for imaging, so-called *dual beam* FIB instruments include an SEM column, which can be used for non-destructive electron imaging. Depending on the manufacturer of the FIB machine, there are differences in the procedure for fabrication of an S-CGH. These differences are associated primarily with the electronics and software that manage beam scanning and patterning. Most of the following discussion is based on the authors' experience with FEI (now Thermo Fisher Scientific) instruments.

After an S-CGH is designed with the aid of a computer and dedicated software, the resulting image can be fed directly to the patterning software that comes with a dual beam machine, or it needs to be converted in a file format that can be read by the software. In the first case, the most common image file formats are .bmp or .png, in the second case vectorial (.dxf or .gdsII) or stream files (.str) are used.
Generally speaking, any file format fed to the software will be used to tell the FIB controller where to position the beam and how long to stay at a certain position. A pixel position in the image is converted to a position in the coordinate system of the beam controller, while the pixel intensity is proportional to the time for which the milling beam spends at that position, *i.e.*, the *dwell time*. This last parameter is what one can use to select between these formats. Most of the afore mentioned image files are 8-bit ones, meaning that the vertical milling resolution in the milling is limited to 256 intensity levels. If higher fidelity in the profile shape is needed, then a different file format is required. This usually translates into the need to use vectorial file formats (.dxf or .gdsII) or a direct coordinate and milling time file format such as stream files (.str), where the resolution in z dimension is no longer a limiting factor.

An additional distinction between image, vectorial and stream files is the order in which the points in the pattern are scanned. For a picture or vectorial format, the FIB pattern handling software allows a choice of scanning direction (*e.g.*, line by line or column by column, in different directions, or spiraling). In addition, all software packages typically allow to choose the number of passes across the sample. The total milling time can, therefore, be subdivided into longer dwell time for fewer passes or shorter dwell time for more passes. These aspects will be covered in detail later in this chapter.

**Optional procedure: Au coating**

In general, when performing an observation of a synthetic hologram using low-angle diffraction, a central spot and additional lateral spots can usually be identified. As outlined above, the central spot in the Fraunhofer plane is referred to as the $0^{th}$ diffraction order, whereas the lateral spots are non-zero diffraction orders that arise from periodicities in the sample. The part of the electron wave function that impinges on the S-CGH, which contains the patterning periodicities, is diffracted and contributes to the intensities of the diffraction spots, *i.e.*, electron holograms encoding the wave function of interest. All parts of the wave function that impinge on unpatterned areas in the surroundings of the S-CGH, along with unscattered electrons and a contribution from non-ideal S-CGH fabrication, will contribute to the intensity of the central spot. In order to minimize the intensity contribution from surrounding unpatterned areas, which can suppress the contribution from the S-CGH, it is possible to first deposit a thick (~150 nm) layer of Au by sputtering or evaporation, followed by FIB removal of this Au layer only in the area where the S-CGH will be fabricated. Although this procedure is rapid and straightforward, the downside is an increase in SiN surface roughness due to the roughness of the Au coverage, which is subsequently projected onto the SiN surface after Au removal by FIB milling. An alternative Au coating procedure using EBL, which is less straightforward and more time-consuming, preserves the initial SiN surface roughness but may leave residues.

*2.2 FIB milling calibration*

After the file format is chosen, the milling process requires calibration to be able to mill reproducible patterns with a well-defined groove depth and, in the case of phase S-CGHs, to obtain a desired phase change. Several factors play a role in determining the milling yield, including the beam current, dwell time and number of passes. The calibration process requires to produce a series of simple patterns, controlled by one of the source files described above, each of which has the same size (*i.e.*, number of pixels) but different milling times and different real pixel sizes. For example, it is possible to use a square-shaped pattern, in which half of the side is milled twice as much the other half, such as that show in Fig. 15.

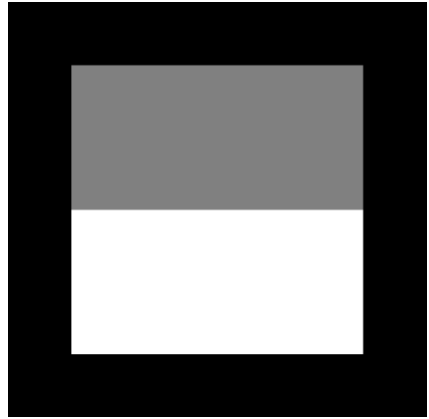

*Figure 15 - Representative calibration pattern image. White corresponds to a pixel of maximum intensity, i.e., to the longest dwell time, while black corresponds to "zero" intensity, i.e., to no milling. Grey corresponds to an intermediate intensity (i.e., dwell time).*

The next step involves reproducing the pattern on a membrane of known thickness. The milling time defined by the user should be quite long since the process has to be manually stopped once one side of the pattern (the white one in this case) completely breaks (as in Fig. 16(d)). A clear sign that the membrane is about to break is the appearance holes as in Fig.16(c). By knowing the total pattern size, the beam current, the milling time and the physical size of each pixel, it is possible to determine the dose and therefore the milling rate for that specific pixel size and current. Figure 16 shows an example of such a procedure, in which the total milling time is increased by changing the number of repetitions while keeping the pixel dwell time at $10^{-4}s$, illustrating (b) slight bending, (c) local milling through and (d) severe milling through the membrane. In this way, the milling depth can be measured and an estimate of the milling rate can be obtained.

Another critical parameter is the pixel size in the image or stream file. The pixel size is the area every pixel from the image will occupy on the substrate, and it is equivalent to the square of the distance between neighboring pixels. The pixel area can be varied in many ways. For instance, instruments controlled by a Raith scan and control unit typically allow the user to choose the pixel size once the pattern image is loaded, to modify the pattern and to impose custom sampling conditions and milling mode on the designed pattern. . Instruments from Thermo Fisher Scientific instead define the pixel position in the imaging reference frame. Therefore, for the same pattern, the choice of magnification leads to different pixel sizes.

This information can be used as a starting point for pattern milling aimed at achieving a desired phase shift. Greater accuracy in the calibration may be achieved by repeating the procedure using different pixel sizes and ion beam currents. The calibration should, in principle, be valid while the ion beam aperture, which defines the beam current and spot size, remains unchanged.

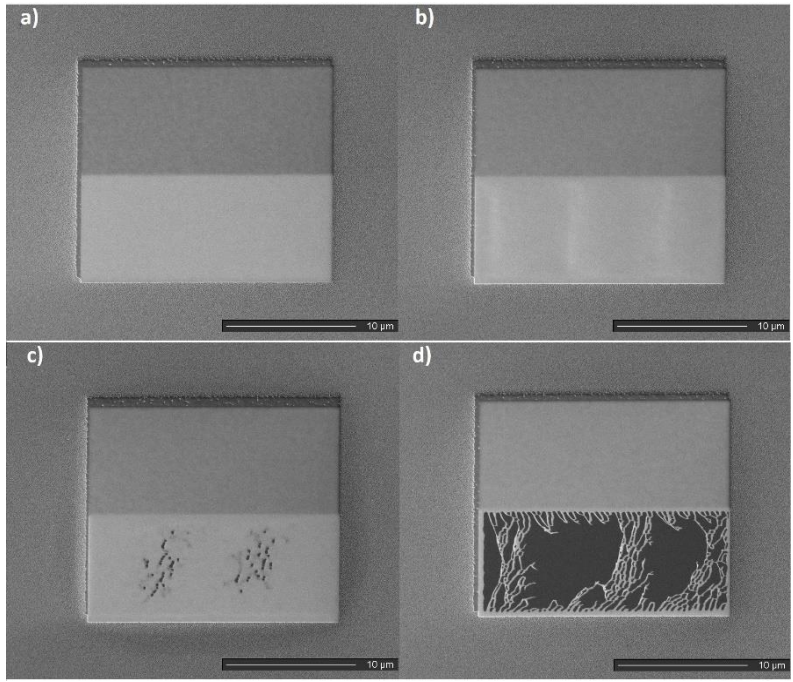

Figure 16: SEM images illustrating fabrication of the pattern shown in Fig. 15 on a $Si_3N_4$ membrane. From a) to d), the total milling time and number of repetitions were increased linearly. The pixel size was ~ 30 nm, the ion current was ~ 104 pA and the accelerating voltage was 30 kV. A low electron energy was used to enhance surface sensitivity during imaging.

This method is effective for determining the milling rate. By increasing the number of tests, it is possible to decrease the error statistically. As a rule of thumb, we repeat the procedure 4 to 6 times for each ion current that will be used for patterning.

Apart from ordinary surface profilometry using methods such as atomic force microscopy, complementary TEM measurements can help to improve the fabrication depth accuracy and to examine if a fabricated S-CGH works appropriately. These methods include low-angle diffracton (LAD), energy-filtered TEM (EFTEM) and low magnification off-axis electron holography. Whereas LAD is available on most modern TEMs and can easily be used to achieve camera lengths of 1.4 km, EFTEM and low magnification off-axis electron holography are less commonly used. The first method requires an energy filter. The second method requires a biprism and the use of free lens control, which can damage the biprism if it is performed carelessly.

By fabricating a series of diffraction gratings that are identical to each other apart from the overall milling time, it is possible to estimate the correct milling time by comparing the diffraction intensity using LAD. For example, for an S-CGH with a sinusoidal modulation, the intensities of the central spot and the first order diffraction peak will depend on whether it has been properly milled. It is good practice to start with larger variations in milling time to be able to assess a wide range of parameters. Subsequently, the process should be refined using a smaller range of parameters. As a rule of thumb, such a process needs two to three iterations to find the best milling time and is therefore time-consuming. Before a good calibration is achieved, at least seven to ten patterns need to be optimized by changing the pixel size or milling current from one run to another. Furthermore, care should be taken to avoid a 2π ambiguity in the fabrication of a phase S-CGH when a large range of thickness values is explored.

EFTEM mapping is a complementary technique, which can be used to provide real space thickness information about the pattern. This technique exploits inelastic interactions between incident electrons and the sample, with scattered electrons losing a small amount of energy that can be measured using an energy filter. The proportion of electrons that have undergone inelastic scattering compared to electrons that have undergone elastic scattering or any scattering at all can yield a value proportional to the local thickness by using the log-ratio method [10]. This value, multiplied by the electron mean free path, provides the local thickness, which can be compared to the intended thickness. In this way, it is possible to reconstruct an x-y map with additional thickness information. As the thickness determines the phase shift, it is possible to use the resulting thickness map in computer simulations of electron beam propagation to understand how the hologram's phase and amplitude information influence the details of LAD patterns.

The use of low magnification off-axis electron holography as an alternative method to validate the quality of a S-CGH and to calibrate the FIB machine requires setting up the TEM in a non-standard configuration. This technique allows the phase and amplitude of a large region of interest of a sample to be measured directly. The region of interest is usually limited to 30 μm × 30 μm and the approach requires milling of a large window near the S-CGH for the reference wave. A linear gradient may need to be removed from the recorded phase image during post-processing.

It is necessary to point out that the calibration process needs to be repeated every time the substrate material for the S-CGH is changed. If more complex patterns are required, then the methods can provide valuable information for their fabrication. It is advisable to carry out a new calibration for every new pattern or experimental condition if high accuracy is required.

Figure 17 summarizes the fabrication process of an S-CGH using FIB milling. It also provides an intuitive recipe for calibration of the FIB machine. For simplicity and clarity, only the main steps are shown.

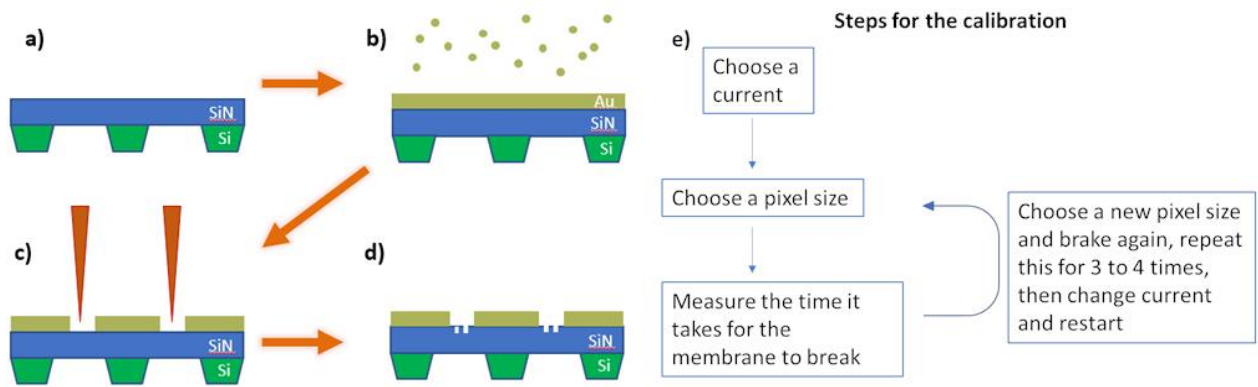

*Figure 17: Schematic diagrams showing the typical steps in the fabrication of an S-CGH using FIB milling: (a) Fresh device; (b) Au evaporation; (c) Au removal and FIB patterning; (d) Grooves in the membranes; (e) Simple algorithm for the calibration process.*

*2.3 Optimization of FIB milling pattern reproducibility*

Once the milling process has been calibrated, it is possible to start S-CGH fabrication. The calibration process focuses on estimating the milling rate of the FIB instrument, while the optimization process is used to fine tune the parameters to achieve an optimal result. Parameters that can be optimized include beam current, pixel size, distance and dwell time, the number of passes or repetitions of the pattern and the scanning strategy. Even the membrane thickness before S-CGH milling will influence the result. This section contains some tips and tricks.

*Optimization of ion current*

The choice of ion current is related to the choice of ion probe size, which ultimately defines the hologram resolution. The primary parameters that should be considered are the total milling time and the pixel size. The pixel size is related to the intrinsic resolution of the S-CGH, with finer details in the profile requiring a smaller pixel size. In general, a higher pattern resolution is desirable. However, there is a limit to how small the pixel size can be, since a resolution that is too high or hologram area that is too large can result in a file whose size cannot be handled by the patterning software, while a pattern resolution that is higher than the milling resolution will not be reproduced properly in the S-CGH.

A Ga ion source on a high-end instrument, at the lowest current, can have a spot size of approximately 5 nm or less. Although the size scales as the square root of the current, the patterning resolution also depends on other factors, such as the local milling time or instabilities, resulting in a larger effective spot size. A lower current is needed for higher resolution, at the cost of a longer patterning time as the sputtering rate depends on the current. However, long continuous patterning times have a higher probability that a drift of the stage or a beam defocus may occur. Although these effects can be reduced by using machines with interferometric stages and higher beam stability, normally a trade-off between ion current and total patterning time must be found. As a rule of thumb, patterning times longer than two hours are not recommended. For these reasons, the current should be chosen carefully to achieve the best resolution for a reasonable patterning time.

***Optimization of dwell time, repetition number, pixel distance and scan direction***

The local milling time, or dwell time, is one of the parameters that can be optimized alongside the pixel distance (if available), number of repetitions and patterning strategy or scan direction. The dwell time influences the final shape of a milled pattern. Figure 18 shows an example of a box pattern, which illustrates the difference between using short dwell times with many repetitions and long dwell times with few repetitions, for the same total dose. The former approach (Fig. 18a) results in a rectangular box profile with mild redeposition on the sidewalls, while the latter approach (Fig. 18b) results in a sloped profile with redeposition effects along the horizontal direction of the serpentine scan [29].

The use of too many repetitions can also be detrimental. Drift of few nm can occur during the "homing" phase at the end of a repetition, though rarely, a small drift of few nanometers can occur, leading to smearing of the end result. A trade-off between the number of repetitions and the dwell time is required, while avoiding the use of long dwell times and large numbers of repetitions.

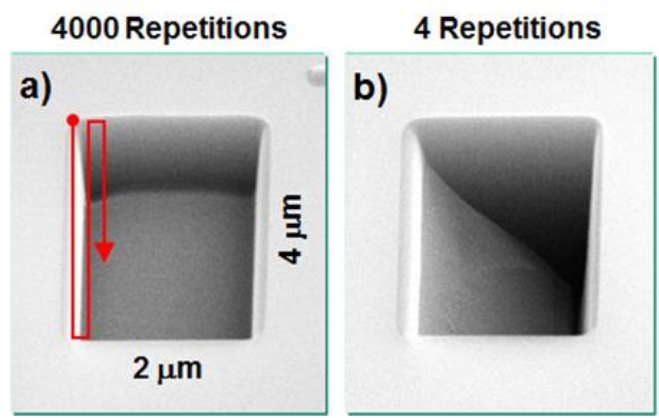

*Figure 18: SEM image (tilted view) of a box pattern milled using: (a) A short dwell time and many repetitions; (b) A long dwell time and few repetitions. The serpentine beam scan is shown using a red line.*

The pixel-to-pixel distance can determine the amount by which adjacent pixels overlap. Clearly, the use of a very large pixel-to-pixel distance (*i.e.*, a highly negative overlap) is detrimental, as the end result is a dotted pattern. Conversely, the use of a very short pixel-to-pixel distance increases the patterning time and file size. A 0 to 50% pixel overlap is ideal in the production of S-CGHs, however pixel-to-pixel distance doesn't affect as much the final resolution of the S-CGH as other factors (primarily the ion current, *i.e.*, the probe size).

The "scanning strategy" determines the path that the beam follows. The most common approach involves the use of zig-zag scanning, as shown in Figs 19a and 19b. An alternative approach involves spiral patterning, as shown in Fig. 19c [30]. It is important to use the best possible scanning strategy because the scanning direction and path contribute to determining where material is redeposited. For zig-zag scanning, redeposition is mainly found on the opposite side to the scanning direction, as shown in Fig. 18b. If long rows are being patterned, it is then suggested to scan the beam along the rows instead of perpendicular to them. For spiral scanning, the continuous "back and forth" motion should allow for a "cleaner" result. However, few examples have yet been presented in the literature [30].

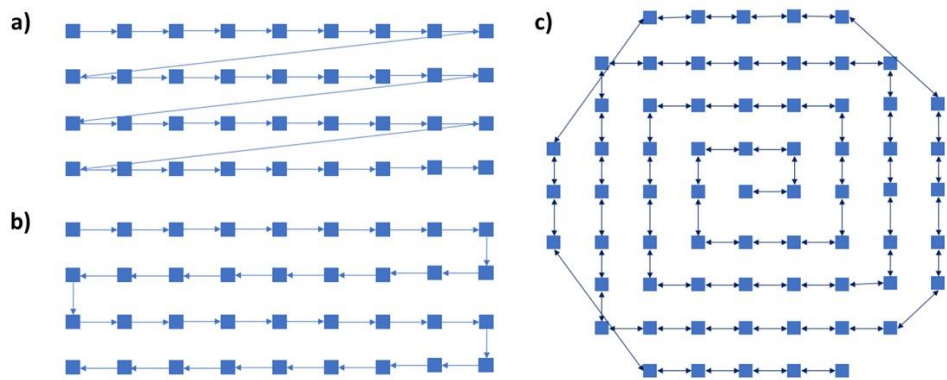

*Figure 19: Examples of patterning strategies. The distance between symbolic pixels has been increased for ease of visualisation.*

*2.4 EBL for S-CGH*

EBL usually requires a series of steps and controlled processes to achieve a final result, but can be used to produce features as small as a few nm and to mass-produce S-CGHs. The typical workflow for production of an S-CGH is shown in Fig. 20. In this case, a negative resist is used. It is also possible to use a positive resist together with reactive ion etching to transfer the pattern, however, it usually leads to poorer results.

Calibration procedures are also needed for EBL. These are less time-consuming when using a negative resist such as hydrogen silsesquioxane (HSQ), which polymerizes into $SiO_x$ when illuminated by an electron beam and has a mean inner potential similar to that of $Si_3N_4$. A standard procedure for selecting the dose involves creating a dose matrix of small features of the pattern that one wants to reproduce. Milling rate does not need to be considered in this case, since the resist thickness dictates the peak-to-valley height.

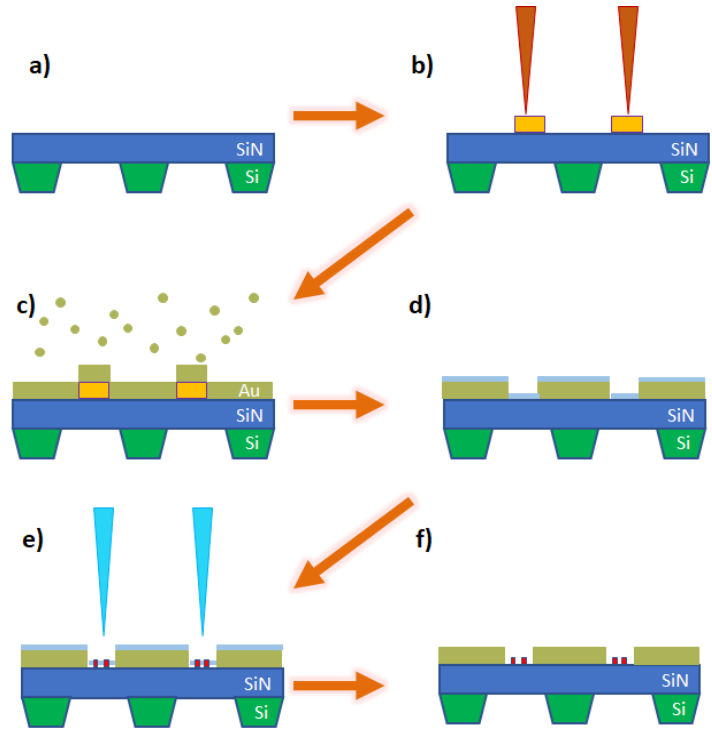

*Figure 20: Schematic diagrams showing: (a) A fresh device; (b) EBL and a developed (negative) resist; (c) Au evaporation; (d) Lift-off and HSQ spin coating; (e) EBL; (f) Developing the HSQ.*

The steps required for preparing an S-CGHs using EBL can be summarized as follows. First, a layer of negative resist is spin-coated on the TEM membrane, patterned into the desired S-CGH enclosure shape and developed. A layer of Au or any other metal (with a high atomic number) is evaporated onto the device, with the metallic layer used to block a portion of the incoming beam. The use of an adhesion promoter of Cr or Ti is encouraged before depositing the metal of choice. The device is then immersed in a resist remover to achieve lift-off of the metallic layers that were on the previously developed resist, in order to prepare the canvas for the S-CGH. HSQ or another resist of choice can now be spin-coated to a desired thickness, patterned and developed. At this point, a few nm of metal or amorphous C can be flash-evaporated onto the developed pattern to balance the generation of secondary electrons in the TEM. More details about the fabrication process and the steps and exact parameters can be found in the paper by Mafakheri et al. [31] and many others related to the EBL technique.

Limitations of the EBL technique include the fact that the thickness is fixed, so one needs to fine-tune the spin coating process to achieve the required thickness for the phase shift. In addition, the pattern profile is either squared or sinusoidal and it is difficult to achieve a blazed profile. Most importantly, multiple steps are required to complete the process and the final devices are small and fragile, meaning that they have to be handled carefully during processing. However, the advantages of EBL are manifold. The $Si_3N_4$ membrane thickness can be reduced to only 15 nm as it is only a supporting layer, whereas for FIB milling it is normally at least 75-100 nm before patterning. The use of a thinner membrane reduces inelastic scattering, background noise and absorption. It also allows the use of a lower electron dose during patterning and results in the generation of fewer secondary electrons in the resist-supporting substrate, opening up the possibility to achieve sub-10-nm-sized features if the process is well optimized.

Even for EBL-fabricated S-CGHs, it is possible to adjust the fabrication procedure to obtain finer details. As a result of the large number of steps, a tedious process of trial and error may be required. Examples of possible improvements include:

- Changing the pre-patterning baking temperature or adding a post-pattern baking step;
- Searching for the proper dose and using proximity correction;
- Adjusting the development temperature and time, as some resists provide higher contrast when they are developed at a lower temperature for longer than at room temperature [32], while others behave in a similar manner when developed at higher temperature [33];
- Developing an understanding of the chemistry of the resist to find an optimal developer;
- Test different thicknesses of silicon nitride or other supporting layers.

Some of the inherent limitations of EBL and FIB milling have recently been overcome by using a thermal scanning probe instead of an electron probe for patterning, resulting in higher accuracy and greater control in patterning depth and morphology [34].

**2.5 Experimental limitations of the use of synthetic holograms in microscopy**

The use of S-CGHs can be effective for the realization of complicated phase patterns for wave front control. However, their primary drawback is that they are static. Their exchange with a different one in the aperture plane of the microscope usually requires breaking of the vacuum of the microscope column. Alternative approaches, such as the use of multipoles of spherical aberrations correctors [35], electrostatic fields [9] or programmable phase plates [36] are still far from reaching the same level of arbitrary wave shaping with a similar number of pixels. Thin synthetic holograms are therefore still preferred for many experiments where a well-known effect is sought for, despite the fact that they require the insertion of additional material in the electron beam path, which can result in: 1) Inelastic scattering and decoherence; 2) A reduction in beam intensity; 3) Contamination, damage and aging of the device as a result of electron beam exposure; 4) Charging of the device during operation.

It should also be noted that the use of thin membranes as patterning media for S-CGHs typically suffers from local thickness variations on a scale of a few nm, resulting in a "frosted glass" effect that is similar to the effect on light crossing a turbulent or inhomogeneous medium. Even the elastically scattered part of the electron beam will therefore have a lateral spread in momentum due to the membrane. Furthermore, different forms of inelastic scattering will reduce the beam current and increase the lateral distribution.

Over time, the beam alters the groove profile from the desired phase profile. This effect is more significant if the synthetic hologram is in the condenser plane, where the electron beam current is higher. Experimentally, the quality of a synthetic hologram is found to deteriorate quickly due to contamination (local C deposition can form in only a couple of days). In contrast, damage (*e.g.*, from knock-on effects and irradiation) tends to be slower, with minor profile alterations becoming apparent after one week of intensive use. It is therefore important to take care of vacuum quality in the TEM column and to be careful during operations such as sample exchange to decrease the probability of contamination. It is also important to avoid concentrating the electron beam to a spot on the S-CGH during any phase of operation. The most serious problem is potentially charging, in particular because SiN is an insulating material, from which it can be difficult to dissipate charge generated by the electron beam. As mentioned above, most of the membrane onto which the S-CGH is patterned is covered by a relatively thick Au layer that allows to dissipate the charge and the electrons only pass through the transparent area of the S-CGH. Whereas the Au layer is efficient in removing charge and partially blocking the beam, the problem can persist in the uncovered area. Experimentally, in the steady state, synthetic holograms are often found to develop a charge density distribution that results in an approximately parabolic projected potential profile, which in turn adds a focusing effect to the hologram phase. It is possible to compensate for such an effect by using the microscope lenses. However, the required compensation can depend on the electron dose, i.e., the higher the dose, the greater the effect. Furthermore, when using large synthetic holograms and unfavorable materials such as HSQ, a steady state is sometimes never reached and the additional phase contribution may vary over time. Possible solutions to this problem include the use of more conductive materials such as C, or coating both surfaces of the S-CGH with a thin layer of metal or C. The use of thinner synthetic holograms is also helpful. An alternative approach involves using amplitude S-CGHs, which are virtually all conductive [37]. As the presence of thin material bridges exposed to vacuum makes such structures mechanically unstable and difficult to fabricate, the structure can be strengthened by substituting the separate lines with a cross-grating. The diffraction orders are then dispersed in two directions, with an overall reduction in the efficiency of the order of interest and greater difficulty in isolating it, as shown in Fig. 21.

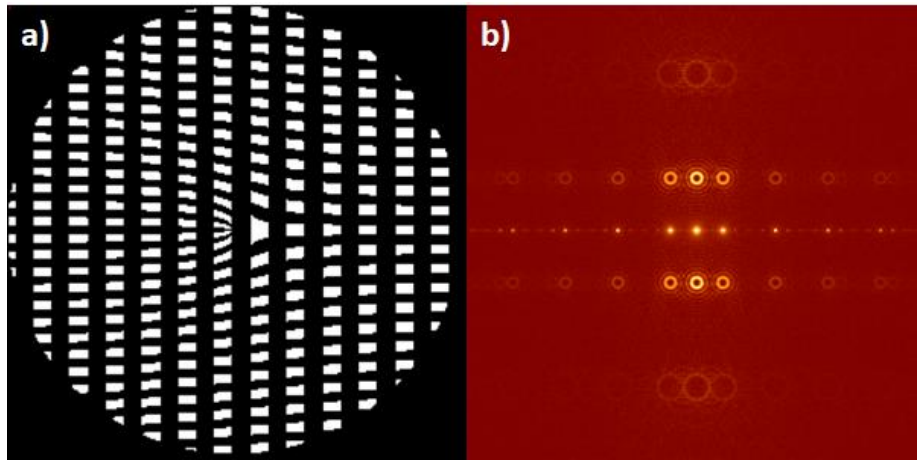

*Figure 21: a) Example of an amplitude S-CGH with a grid-like structure for improved mechanical stability; b) The resulting diffraction pattern, which forms a two-dimensional array of beams [36].*

# Chapter 3 - Examples

## 3.1 Phase S-CGH design for the generation of electron beam vortices carrying orbital angular momentum

The generation of electron vortex beams (EVBs) was first demonstrated in 2009 and 2010 by three groups. The approaches involved using "spiral phase plates" constructed from thin films of graphite [38] and S-CGHs with pitchfork designs [39, 26]. In some of the first experiments, EVBs were generated using amplitude S-CGHs or similar structures. Since then, most research groups have used phase or mixed amplitude-phase S-CGHs, which have higher efficiencies. New methods for the generation of EVBs have been presented [40, 41, 42] and the topic has matured sufficiently that most efforts are directed towards the measurement of OAM values and increasing applications in the fields of plasmonics, studies of magnetic materials and chiral structures such as proteins. In a circular symmetrical reference system, an EVB has an angular-dependent helical phase term, which can be described by the expression:

$$\varphi(l,\theta) = \ell\theta \, , \tag{47}$$

where $\ell$ is the OAM eigenvalue of the Schrödinger equation solved in cylindrical coordinate (also known as the topological charge or OAM quantum number) and $\theta$ is the angular coordinate. The wave function of a generic EVB is then given by the expression

$$\Psi_{helical} = A_0 e^{i\ell\theta} \, . \tag{48}$$

Some of the most prominent strategies for creating EVBs are described below. Further details about EVBs and vortex beams in general can be found elsewhere [37, 43, 44, 45, 46].

### 3.1.1 Spiral design

The simplest way to generate an EVB using a S-CGH is to design an in-line [17, 47] phase S-CGH that has a spiral/ helical form, similar to that shown in Fig. 1, in which a smoothly-varying thickness profile is used to tune the phase shift imprinted on the wave front of the outgoing beam.

This design, in its simplest form, is an inline S-CGH and its realization requires good control of the fabrication process for the reasons outlined in Chapters 1 and 2. However, a well calibrated machine makes fabrication straightforward. An EVB with topological charge $\ell$ can be generated using a spiral phase plate in which the total phase shift over a complete revolution is $\Delta\varphi = \ell \cdot 2\pi$.

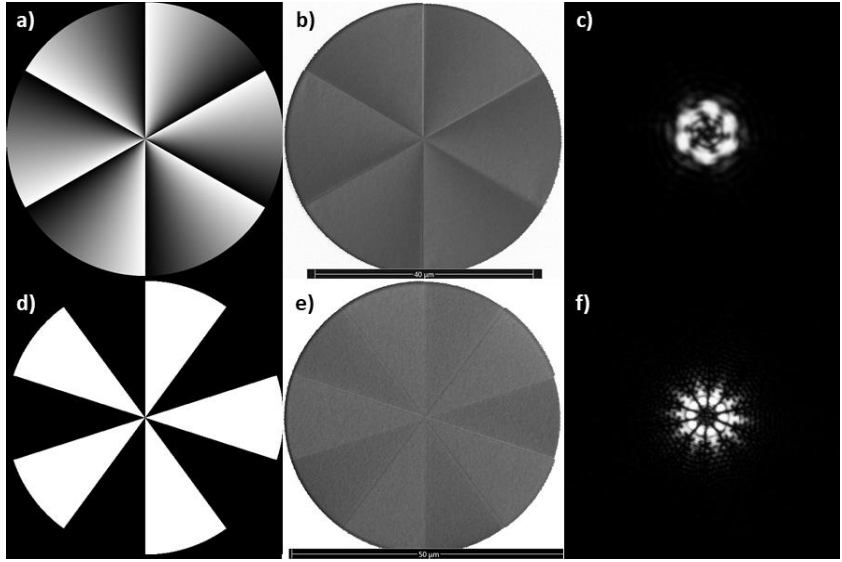

*Figure 22: a) Phase of an EVB used to fabricate a phase-S-CGH with a spiral/ helical design for EVB generation for l = 6ℏ. The phase varies from 0 (black) to $2\pi$ (white) and goes from 0 to $12\pi$ over a complete revolution. (b) SEM image of a phase S-CGH corresponding to a). c) Experimental EVB in the Fraunhofer plane. d) Phase and e) phase S-CGH for an EVB with $\ell = \pm 5\hbar$. f) Experimental EVB.*

A typical design of an EVB with a spiral phase is shown in Fig. 22a, with phase ramps in six angular sections, in each of which the phase shift goes from 0 to $2\pi$. The outgoing EVB therefore carries an OAM value corresponding to $\ell$ = 6.

This design allows a superposition of EVBs to be generated. A beam that is generated from two superimposed and has no azimuthal current is referred to as a "petal beam". For example, a phase S-CGH can be used to generate an electron beam corresponding to a coherent superposition of $\ell=-5$ and $\ell =+5$ by summing the wave functions for EVBs with $\ell = 5$ and $\ell = -5$ and calculating the phase of the resulting wave function. Mathematically, the phase is $\Delta\varphi = arg\,(sin\,(l\,\theta))$, corresponding to alternating values of 0 and $\pi$. Figure 22d shows the phase of a beam that carries an OAM superposition of $\ell=\pm 5$, with white corresponding to a phase shift of $\pi$ with respect to black areas. For a generic EVB generator with a spiral design, the enclosure is a circle just as for a conventional aperture and the physical dimension is typically $10 - 50\ \mu m$.

### 3.1.2 Pitchfork design

A pitchfork design can be used for both amplitude [38] and phase [31] off-axis S-CGHs, with EVBs generated in the $n^{th}$ diffraction order, where *n* can vary from 1 to infinity. The design is based on an interference pattern between a plane wave $\Psi = A_0 e^{i(k_x x + k_z z)}$ and a helical wave in the $z = 0$ plane, in the form

$$I = 2|A_0|^2(1 + \cos(k_x x - \ell\theta)) \,, \tag{49}$$

from which it is possible to find the phase term of the interference wave to design the pitchfork S-CGH. As described in section 1.4.1, in order to generate a pitchfork S-CGH the argument of the profile function is

$$\alpha(x, y) = \ell\xi + 2\pi x \,, \tag{50}$$

where $x$ is one of the two in-plane coordinates, $\xi = ArcTan\left(\frac{y}{x}\right)$ and $\ell$ is the topological charge. The planar Cartesian coordinates $x$ and $y$ are expressed in units of the grating spatial period $\Lambda$.

Figure 23 shows the bi-dimensional profile functions $f(\alpha)$ for a pitchfork design with $\ell = 2$. Each pattern is obtained by combining the generic profile functions described in section 1.4.3 and Eq. 50, such that:

- $f_{sqrd}(\alpha) = \frac{1}{2}(1 + Sign(sin(\ell\xi + 2\pi x)))$;  [Fig.23a]
- $f_{cos}(\alpha) = \frac{1}{2}(1 + cos(\ell\xi + 2\pi x))$;  [Fig.23b]
- $f_{trian}(\alpha) = \frac{1}{\pi}(Sign(sin(\ell\xi + 2\pi x)))(\pi - Mod(\ell\xi + 2\pi x, 2\pi))$;  [Fig.23c]
- $f_{blzd}(\alpha) = \frac{1}{2\pi}(Mod(\ell\xi + 2\pi x, 2\pi))$.  [Fig.23d]

This design is versatile, as it can be used to generate both low-OAM and high-OAM EVBs. However, in the latter case the features in the central part may be so small (in some cases even smaller than a pixel) that they are almost impossible to reproduce using either fabrication technique discussed above. A common strategy involves masking out the central part up to a chosen radius. Although such a mask reduces the transmitted efficiency, an EVB with the correct OAM value is generated [31].

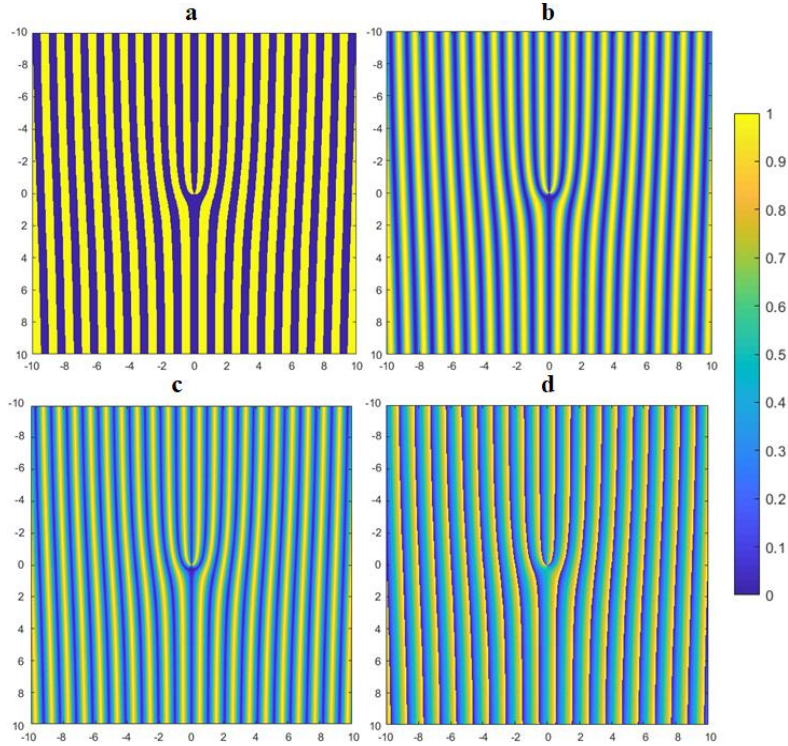

Figure 23: Designs of a pitchfork S-CGH with $\ell = 2$ for a) squared, b) cosinusoidal, c) triangular and d) blazed designs. The color bar represents the value of $f(\alpha)$ at the position of each pixel.

***3.1.3 Case study: Optimization and understanding of a blazed phase S-CGH with a pitchfork design***

In recent years, we have worked on optimizing the fabrication process of a blazed phase S-CGH using FIB milling, in particular for a pitchfork with $\ell = 1$ [48]. We have aimed at reaching the highest diffraction efficiency for one of the two first diffraction orders (100%; see Table 4) by converging most of the intensity in the beam carrying the desired amount of OAM.

In order to reduce the number of variables, most parameters were kept constant, with only the number of passes and the maximum dwell time changed to tailor the phase shift and approach 2π. First, the number of passes was varied for rough optimization, then the maximum dwell time was optimized for finer optimization. The parameters that were kept constant and their values are given in Table 5.

| S-CGH diameter | Ion beam current | CGH resolution | Step Size | Magnification |
|---|---|---|---|---|
| $20\ \mu m$ | $\sim 260\ pA$ | $1024 \times 1024\ px$ | 2 | 10400 X |

*Table 5: Patterning parameters that were kept constant during optimization of the fabrication of a blazed phase S-CGH.*

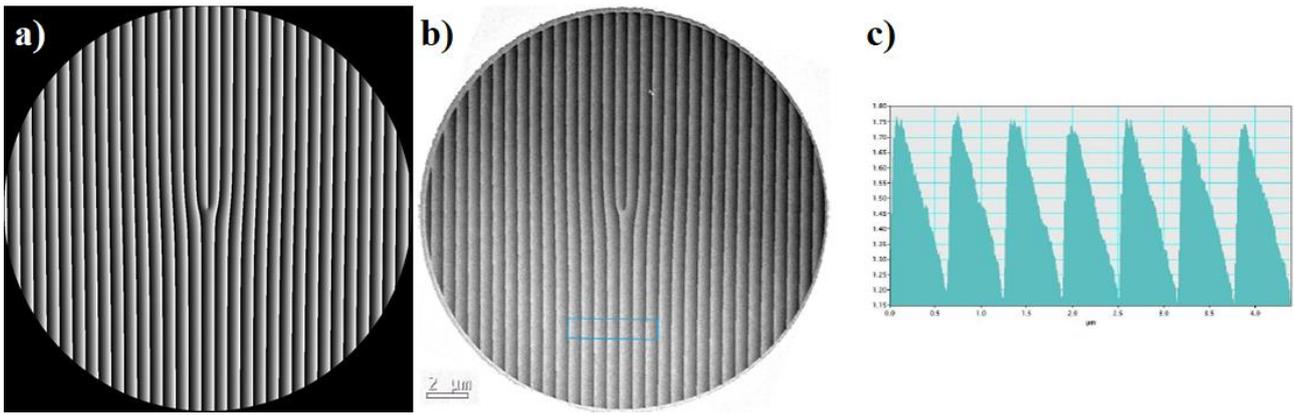

*Figure 24: a) CGH for a blazed $\ell = 1$ pitchfork; b) EFTEM thickness map of the fabricated S-CGH; c) Profile of the region marked by a blue rectangle in b).*

Figure 24 shows the CGH and the best-performing fabricated S-CGH. The patterning parameters for best performance, other than those reported in Table 5, are:

- Number of repetitions: 8 passes.
- Maximum dwell time: 91.6 $\mu s$.

These numbers can vary both between FIB machines and between fabrication sessions, as factors such as laboratory environment, vacuum quality and machine characteristics can influence the fabrication process.

The EFTEM image and line profile in Figs 24b and 24c show that the in-plane periodicity of the pattern is $\sim 600\ nm$ and the distance between peak and valley is $\sim 70\ nm$. This is slightly larger than the required value, which is $\sim 64\ nm$ for 300 keV electrons, as reported in Table 2. The shapes of the peaks approximate the ideal shape of a blazed profile, but differ slightly from one another, with sharp troughs but blunter peaks These effects show some of the limitations of using FIB milling and contribute to the measured reduction in diffraction efficiency. Figure 25 shows that the best-performing sample was able to achieve 66.22% of the transmitted intensity in the +1[st] diffraction order, with the experimental diffraction intensity distributed between the orders in a different manner from that observed in Figs 10 or 13d.

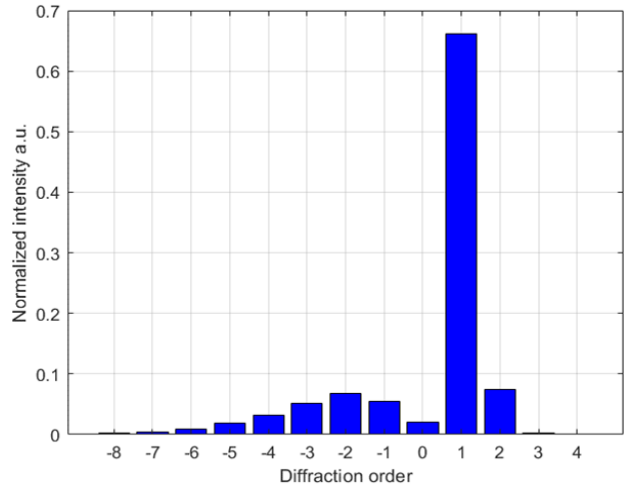

*Figure 25: Experimental distribution of diffraction intensities in the best-performing sample, with the total intensity normalized to unity.*

We used simulations to assess the origin of this behavior. First, we examined the effect of a non-ideal peak-to-valley phase difference by recalculating the intensity distribution for a $\pm 10\%$ phase mismatch from an ideal phase S-CGH. Figure 26 shows that even a 10% mismatch has almost a negligible influence on the diffraction intensity distribution, suggesting that the intensity distribution measured experimentally has a different origin. Although absorption affects the diffraction intensity, as shown in section 1.4.3 and Fig. 10, it mainly decreases the total transmitted intensity, redistributing it almost evenly between the orders.

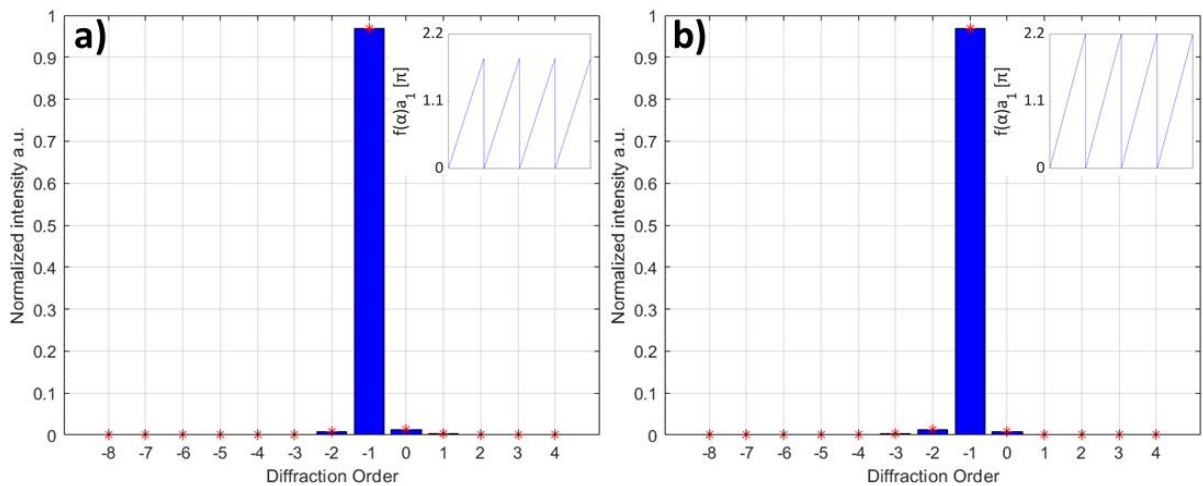

*Figure 26: Diffraction intensity distributions for phase mismatches of a) $-10\%$ and b) $+10\%$. The insets show the corresponding groove profiles.*

Even by considering the effect of both the absorption and the phase mismatch it is still impossible to reproduce the same intensity distribution. We then focused on the profile shape of the S-CGH. Figure 24c shows that the actual shape is closer to a scalene triangle than to a blazed one. The scalene triangular profile function is

$$g(\alpha) = \begin{cases} Mod\left(\frac{1}{s}\alpha(\vec{\rho}), 2\pi\right) & for\ \alpha(\vec{\rho}) < s \\ 1 + \frac{s}{(2\pi - s)} - Mod\left(\frac{\alpha(\vec{\rho})}{2\pi - s}, 2\pi\right) & for\ s \leq \alpha(\vec{\rho}) < 2\pi \end{cases}. \tag{51}$$

This profile function is normalized between 0 and 1 and has its maximum for $\alpha(\vec{\rho}) = s$. The further s is from 0, the more it differs from an ideal blazed profile. Figure 27 shows the intensity distribution for $s = 1.1$. Although the shape difference was accentuated by choosing a high value of $s$, it is likely to be imperfections in the profile shape, including small differences between the shapes of adjacent "teeth", that lead to spreading of the diffraction intensity between the orders.

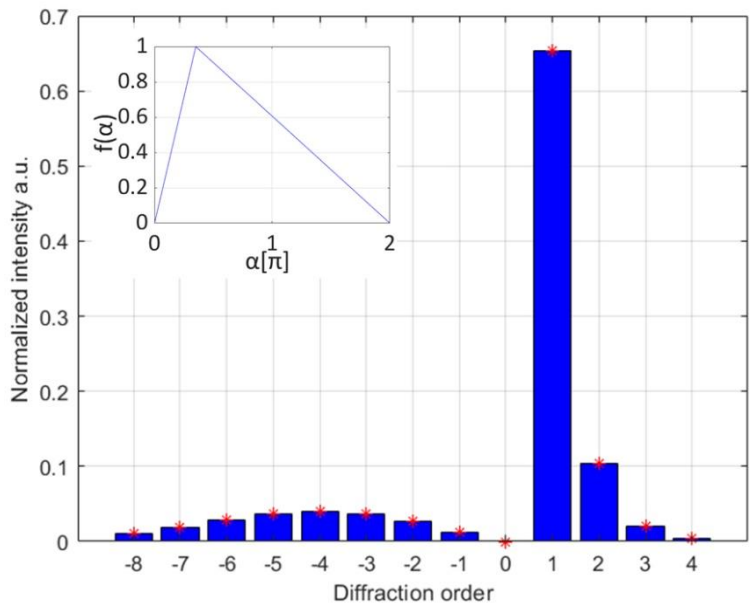

*Figure 27: Diffraction intensity distribution for a scalene triangular profile for s=1.1. The inset shows a groove profile for 1 period.*

In summary, we have been able to model and discover the main factors that limit the operation of a real blazed phase S-CGH. Most of them are related to limitations of the fabrication process. Imperfections of the shape and the phase mismatch usually result from instabilities of the FIB machine. Optimization of the lateral resolution of the FIB machine is likely to bring the greatest improvement. Even a stage shift of only a few nm (for thermal or mechanical reasons) during the fabrication procedure can compromise the result. It may be possible to reduce some of the limitations by using a FIB machine that is designed for S-CGH production or by rethinking the fabrication steps. For example, it has been shown that gas-assisted FIB milling can improve the reproduction fidelity and patterning speed of blazed profiles [49]. However, some effects that arise from inelastic and diffuse scattering, including absorption and background noise, will always be present.

**3.1.4 Generation of EVBs using Gaussian beams**

The vortex beam generators that were described above are characterized by a hard aperture in the hologram plane. The resulting beams are sometimes referred to as "hypergeometric beams" [50]. In light optics, a more suitable class of vortex beams has been derived based on a member of the Gaussian beam family: so-called Laguerre-Gaussian (LG) beams. Exact Gaussian beams are characterized by flat phase wave fronts at $z = 0$ and well-defined amplitude structures, with planes perpendicular to the optical axis that are equiphase surfaces. An in-depth mathematical description can be found in the book by Guenther [51]. An important parameter is the Gouy phase term, which is related to the transverse confinement of the beam and introduces anomalous behaviour in the phase of the beam when it passes through focus [52, 53, 54, 55]. In a TEM, an exact Gaussian beam or a coherent Gaussian beam cannot be obtained easily. However, a Gaussian-like beam that reproduce the intensity of an exact beam can be obtained by converging the beam.

Laguerre-Gaussian beams are of greater interest than simpler Gaussian beams as they are solutions of the paraxial Helmholtz equation in cylindrical coordinates and are eigenstates of both the Fourier transform operation and OAM. In this way, they form a complete orthonormal basis characterized by two discrete quantum numbers $p$ and $\ell$, where $\ell$ is the azimuthal index or topological charge of OAM and $p$ is a radial index, which defines the $(p + 1)$ radial nodes in the intensity distribution. The wave function of a LG beam has the form [56]

$$\psi_{LG\,\ell}^{p}(\rho,\theta,z,t) = \frac{C_{\ell p} z_R}{\sqrt{z_R^2 + z^2}} \left(\frac{\sqrt{2}\rho}{w(z)}\right)^{|\ell|} L_p^{|\ell|}\left(\frac{2\rho^2}{w^2(z)}\right) exp(i(k_z z + \ell\theta - \omega t)) \times$$
$$\times exp\left(-\frac{\rho^2}{w^2(z)} + ik_z \frac{\rho^2}{2R(z)}\right) exp(-i(2p + |\ell| + 1)\xi(z)) \,, \tag{52}$$

where $L_p^{|\ell|}$ is the generalized Laguerre polynomial [57], $C_{\ell p} = \sqrt{\frac{2^{|\ell|+1} p!}{(\pi(|\ell|+p)!)}}$ is a normalization factor, $w(z) = w_0\sqrt{1 + \left(\frac{z}{z_R}\right)^2}$ is the beam waist radius along the propagation axis $z$, $w_0$ is the beam radius in focus, $z_R = \frac{k_z w_0^2}{2}$ is the Rayleigh range, $\xi(z) = ArcTan\left(\frac{z}{z_R}\right)$ and $R(z) = z\left[1 + \left(\frac{z_R}{z}\right)^2\right]$ is the radius of curvature of the complex wave front.

It is possible to demonstrate that the evolution of this kind of Gaussian beam along the optical axis is related only to the Gouy phase $exp(-i(2p + |\ell| + 1)\xi(z))$ and $w_0$, which makes it diffraction-shape-invariant, evolving only by the scale factor $\sqrt{1 + \left(\frac{z}{z_R}\right)^2}$. This is a weaker condition for diffraction invariance than for Bessel beams, which are described in section 3.3, the difference being that Bessel beams are non-normalizable and therefore not exactly realizable experimentally. A series of simulated LG beams with varying indices are shown in Fig. 28.

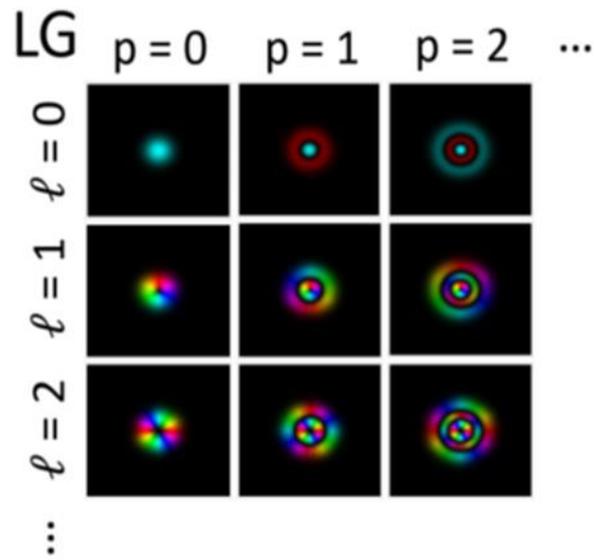

*Figure 28: Examples of Laguerre-Gauss beams with varying indices. The intensities and phase shifts of the wave functions are represented by their brightness and hue, respectively.*

LG beams are of interest to scientists working on magnetic materials and structured waves. For example, a LG wave function is functionally similar to a Landau state wave function [58]. By tuning a LG beam waist, it has been demonstrated experimentally that it is possible to couple them to Landau states [59]. Even though the generated LG beams were not pure, this experimental proof opens the possibility of observing transitions between the states. Furthermore, a LG beam has been used to demonstrate that it is possible to use paired S-CGHs for almost direct phase retrieval of EVBs (and of structured beams in general) in the Fraunhofer plane [60]. Pure LG beams are ideally generated using mixed S-CGHs [61], as described in section 1.5.1. The design and fabrication of mixed S-CGHs are reported in Fig. 29 for two experimental examples of LG beams with different characteristics. The first example (Figs 29a-d) shows a pure $LG_0^{10}$ mode that has a simple circular structure. The second example (Figs 29e-g) shows two states with different OAM and *p* quantum numbers coherently summed together to give a superposition of LG modes with different radial and azimuthal indices. The phase in Fig. 29f, which is the theoretical phase obtained by Fourier transforming the thickness profile of the hologram, illustrates the complexity of the beam. It can be considered as proof of the power of amplitude and phase encoding in a single S-CGH for Laguerre-Gauss beam generation, and in general EVB generation. In Fig. 29d, there are no intensity ripples similar to those present in EVBs generated using a spiral design, as described in section 3.1.1. LG beam generation using different techniques has also been reported [62].

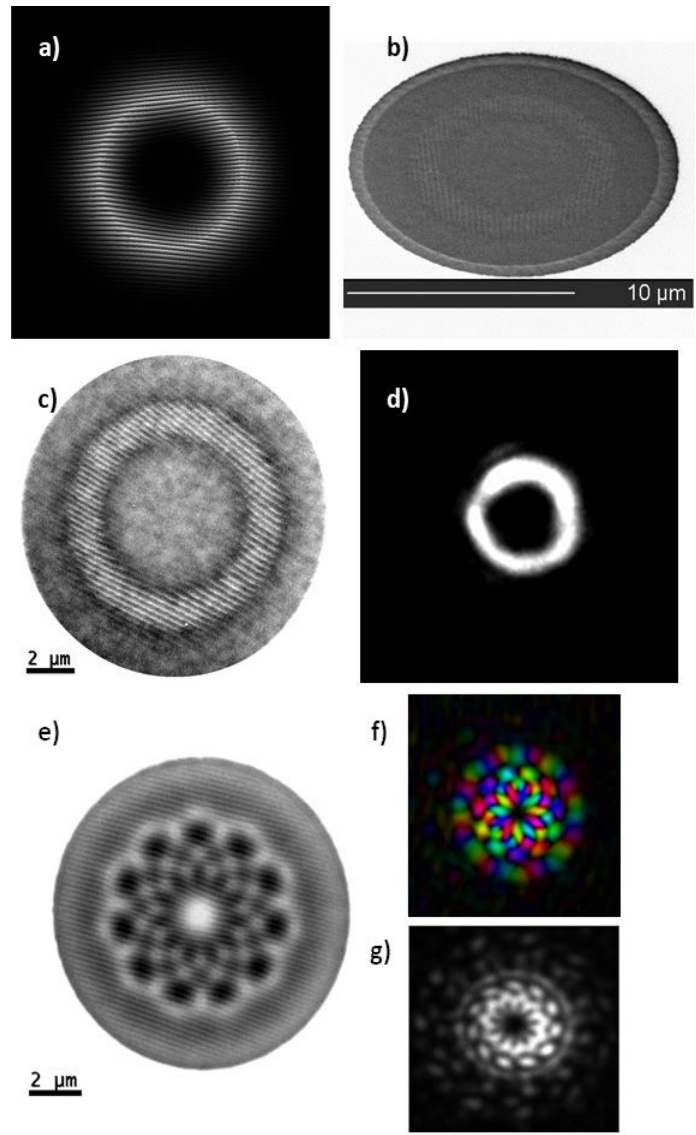

*Figure 29: Steps in the fabrication and validation of two LG beams: a) Phase and amplitude utilized for fabrication; b) Tilted SEM image of the resulting mixed S-CGH (ion current: 300pA; repetitions: 192; magnification: 3900x); c) EFTEM map of the S-CGH and d) diffraction image showing the "donut-like" shape of the generated EVB; e) EFTEM map of superimposed LG beams; f) Simulated amplitude and phase and g) diffraction intensity of the $1^{st}$ diffraction order.*

Whereas LG beams are solutions of the paraxial Helmholtz equation in cylindrical coordinates, Hermite-Gaussian (HG) beams are solutions of the same equation in cartesian coordinates [63]. Although HG beams do not carry OAM, the first vortex beams generated by Allen et al. in 1992 [63] were obtained by using a cylindrical lens to transform high-order HG modes into LG modes. In a TEM, it is possible to reproduce the effect of a cylindrical lens by increasing the astigmatism. This approach has been exploited by Schattschneider *et al.* [64] to measure the OAM of an EVB and to measure the azimuthal (and radial) state for exact LG states.

**3.2 Design and realization of a holographic OAM sorter**

An interesting application of synthetic electron holograms is the development of a device that can be used to measure the OAM spectrum of an electron beam, referred to as an OAM sorter [65]. This device is composed primarily from two S-CGHs: an "unwrapper" S-CGH that unwraps an OAM-carrying electron beam and a "corrector" S-CGH that corrects the phase distortion introduced by the first S-CGH. The incoming electron beam contains the phase information of interest, after having interacted with a sample. The most straightforward example of an OAM-generating sample is the in-line S-CGH described in section 3.1.1. Figure 30 provides a schematic representation of the setup and transformations involved, including OAM generation, unwrapping, correction and detection. In Figure 30, an electron beam impinges on a generator S-CGH and is endowed with a spiraling phase shift with OAM = 1, corresponding to a $2\pi$ phase shift along one complete azimuthal path. The use of an in-line S-CGH simplifies alignment of the beam on the sorter and excludes the effect of tilt (and off-axis aberrations).

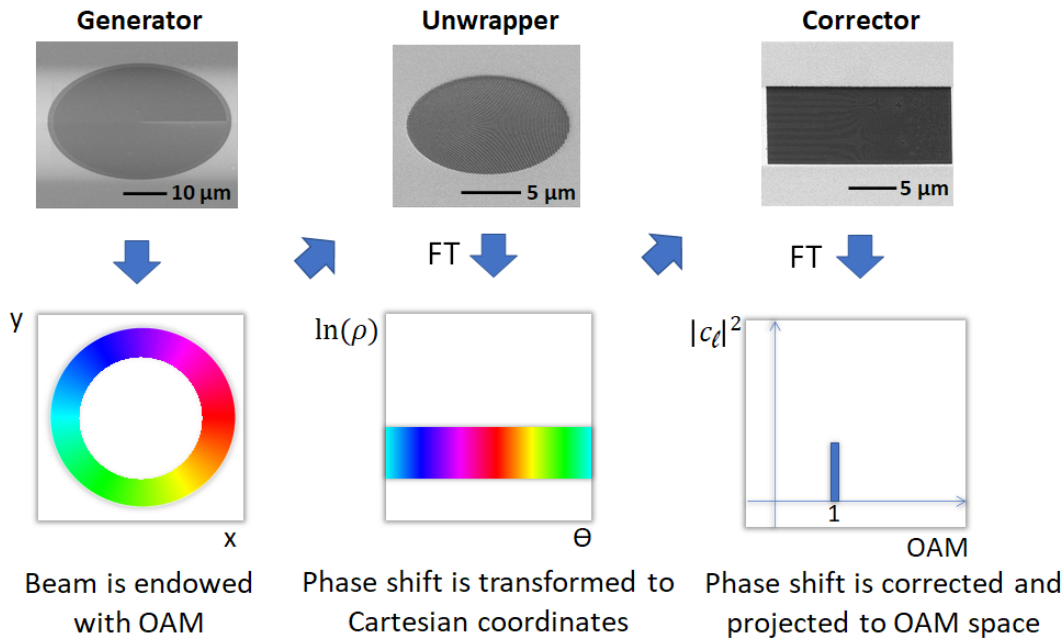

*Figure 30: Sequence of holographic masks and transformations involved in the use of an OAM sorter.*

The electron beam that is carrying OAM is directed onto the unwrapper S-CGH, which performs a conformal transformation from log polar to cartesian coordinates. In this way, the phase information is unwrapped from an azimuthally-varying arrangement to a linear arrangement, so that it is aligned along one cartesian coordinate. The first S-CGH, or Sorter 1 element, is a diffractive hologram. Therefore, the resulting pattern is found in a reciprocal plane. This unwrapping operation introduces a strong phase gradient. After the transformation, this additional phase must be removed. Therefore, the corrector is needed as an additional off-axis S-CGH. The final OAM spectrum is found in reciprocal space. After this correction operation, the OAM value can be found as an intensity spot, whose position from the center of the first diffraction order in reciprocal space is indicative of the magnitude of its OAM value. A calibration procedure using different OAM-generating S-CGHs with known OAM values is required, as each electron-optical configuration can introduce changes in rotation and magnification. Once a device is calibrated, a real sample, which imparts an unknown amount of OAM onto the electron beam, can be studied instead of the generator S-CGH. It is possible to measure the full OAM spectrum of a beam in one acquisition. Applications include measurements of the magnetic moments of dipoles [65], as well as in EMCD [66] and plasmon characterisation [67].

S-CGH fabrication requires a knowledge of the functions that are to be encoded in the unwrapper S-CGH ($\Lambda_1$) and corrector S-CGH ($\Lambda_2$). The phase corresponding to the first element of the sorter is

$$\Lambda_1 = \varphi_0 \, sign\left(sin\left(2\pi a \left| y \arctan\left(\frac{y}{x}\right) + x \, ln\left(\frac{\sqrt{x^2+y^2}}{b}\right) + x \right|\right)\right), \tag{53}$$

where $a$ and $b$ are parameters that are used to optimise the experimental efficiency, while *sign* denotes the sign function. The phase corresponding to the second element of the sorter is

$$\Lambda_2 = \varphi_0 \, sign\left(sin\left(2\pi ab \, exp\left(-2\pi \frac{u}{a}\right) cos\left(2\pi \frac{v}{a}\right)\right) + 2\pi cv\right), \tag{54}$$

where $u = -a \, ln\left(\sqrt{x^2+y^2}/b\right)$, $v = a \, \arctan\left(\frac{y}{x}\right)$ and $c$ is an additional scaling parameter. For the unwrapper shown in Fig. 30, the parameters were $a$ = 2, $b$ = 0.01 and $c$ = 0.6. They can be tuned to match the relative S-CGH and holographic beam sizes. The peak-to-trough depth should maximize the diffraction efficiencies. Although the device can be used in any TEM, the electron-optical configuration is challenging and the use of free lens control and additional sets of lenses and apertures is recommended.

**3.3 Bessel beam**

A third example of a possible application of S-CGHs is the generation of a non-diffractive Bessel beam. Bessel beams were first mathematically modeled by Durnin [68]. Experimentally, they were realized as photon quasi-Bessel beams, an approximation of Bessel beams, which had the same properties over finite distances [69]. Durnin and colleagues defined Bessel beams as beams "whose central maxima are remarkably resistant to the diffractive spreading commonly associated with all wave propagation" [70, 71]. A Bessel beam can be considered as a coherent superposition of conical plane waves, or as a set of plane waves propagating on a cone. Apart from being non-diffractive, they are also "self-healing", so that (apart from an overall decrease in intensity) they can recover their intensity profile. Moreover, a zeroth order Bessel beam has a smaller central spot diameter and longer depth of field than other ordinary beams [72].

In light optics, the generation of Bessel beams, or more precisely quasi-Bessel beams, has been achieved in a number of ways. The simplest approach is to use an annular slit or ring aperture [68]. This method works since the Fourier transform of a Bessel beam is a ring. A more efficient method is to use axicon lenses [73, 74, 75, 76], which remove the on-axis intensity oscillation, resulting in a smooth intensity variation in the beam propagation direction. Other methods are based on S-CGHs [77], SLMs [78, 79] and cavities [80, 81].

In recent years, by taking inspiration from light optics, different methods have been adopted for the generation of electron quasi-Bessel beams. In 2014, Grillo *et al.* reported the use of an S-CGH to generate non-diffractive quasi-Bessel beams that were able to propagate for 0.6 m without noticeable spreading of their central maximum and could reconstruct [82, 83]. Taking inspiration from initial experiments by Durnin, Saitoh and colleagues used annular slits to generate quasi-Bessel beams [84]. In 2017, Zheng and colleagues used magnetic vortices with circular magnetic moment distributions, which are naturally present in soft magnetic thin films, as axicon lenses [85]. A generic Bessel beam wave function can be expressed in the form

$$\psi(\rho,\phi,z;t) = J_n(k_\rho\rho)e^{in\phi}e^{i(k_z z - \omega t)} \;, \tag{55}$$

where $\rho$, $\phi$, $z$ are cylindrical coordinates, $J_n$ is the $n$-th order Bessel function of the first kind, $n$ is an integer, $k_\rho$ and $k_z$ are the transverse and longitudinal components of the wave vector, respectively and $k^2 = k_\rho^2 + k_z^2 = \frac{2m\omega}{\hbar} = (\frac{2\pi}{\lambda_{dB}})^2$, where $m$ is the electron mass, $\hbar$ is the reduced Planck constant and $\lambda_{dB}$ is the electron's de Broglie wavelength. A Bessel wave function is a well-known non-normalizable solution of the Schrödinger equation of a free electron in cylindrical coordinates. From Eq. 55, it is possible to notice that the probability density is independent of both time and $z$, and is equal to $J_n^2(k_\rho\rho)$.

The phase S-CGH used by Grillo *et al.* [82] imprints on the transmitted beam the phase modulation

$$\varphi(\rho,\phi) = \varphi_0 sgn[\cos(k_\rho \rho + n\phi + g\rho \cos\phi)] \ . \tag{56}$$

The resulting off-axis Fresnel hologram has carrier frequency $g = \frac{2\pi}{\Lambda}$, where $\Lambda$ is the grating spatial period. In this formula, the chosen profile shape was a squared one with argument $\alpha(\rho,\phi) = k_\rho \rho + n\phi + g\rho\cos\phi$, where $n$ is the OAM topological charge. The $\alpha(\rho,\phi)$ that was used is similar to that in Eq. 50, *i.e.*, the pitchfork design. The resulting quasi-Bessel beam was an OAM-carrying one. Figure 31 shows a fabricated phase S-CGH for quasi-Bessel-beam generation, the CGH that was used to produce it and an experimental diffraction image, in which it is possible to observe the generated quasi-Bessel beam. The typical dislocation of a pitchfork design is visible at the center of Fig. 31b.

In a later paper [83], by switching to a cosinusoidal profile and optimizing the fabrication procedure, Grillo and colleagues increased the transmission efficiency by $37 \pm 3\%$. They pointed out possible application fields of quasi-Bessel beams generated using S-CGHs: smaller aperture radii are best suited for STEM, while larger radii are best suited for interferometry. Applications of such structured beams in electron include classical techniques such as tomography [86] and strain mapping [87], as well as conventional STEM, low dose STEM and HR-STEM [82, 83, 88, 89].

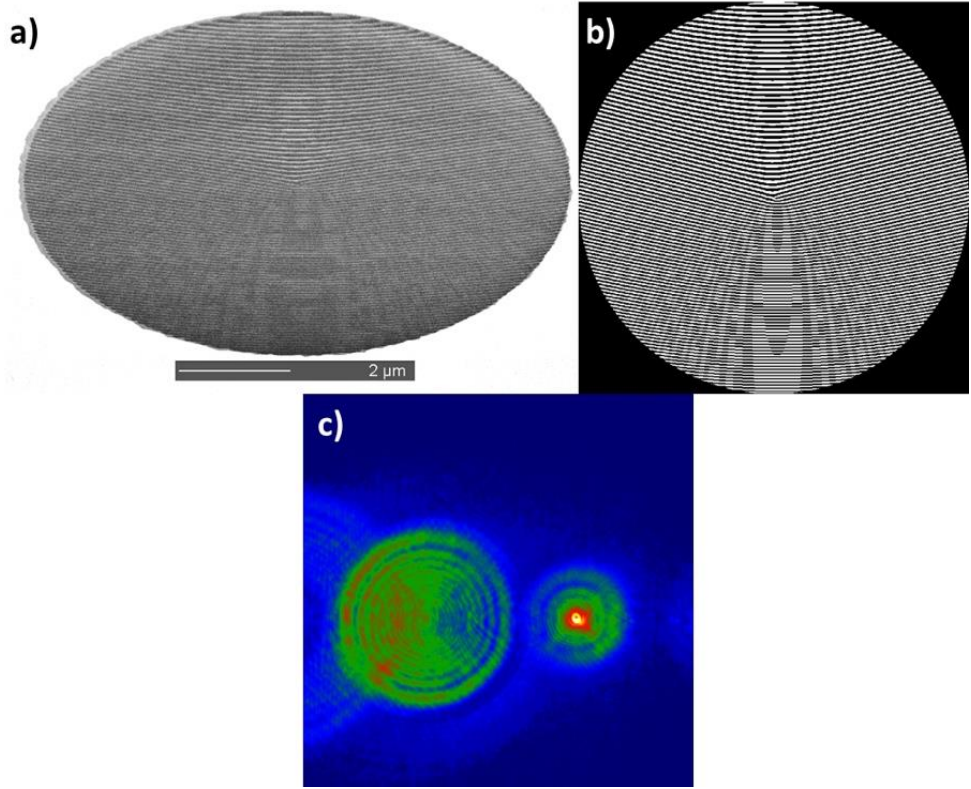

*Figure 31: a) SEM image of a phase S-CGH used for the generation of quasi-Bessel beams with ℓ =2, recorded with the stage tilted to highlight the three-dimensional features. b) Profile function used to obtain*

*a) by FIB milling. c) Experimental diffraction image. The first diffraction order on the right shows a quasi-Bessel beam. The parameters used in [79] were used to create b).*

### 3.4 C$_S$ corrector

A final example of beam shaping does not rely on the scheme of chapter 1.4. A problem that has long affected electron microscopy is the presence of spherical ($C_S$) aberration in any magnetic lens that has cylindrical symmetry. Although a solution has been found by using a complicated set of multipoles [90, 91], it is interesting to determine whether one can produce an S-CGH that is able to compensate for spherical aberration by introducing, in the condenser aperture plane, an equal phase of opposite sign to that of the $C_S$ aberration. The desired phase $\alpha$ is not known in the diffraction plane, but directly in the S-CGH plane. Therefore, the aim is to correct for the $C_S$ aberration in a STEM probe by using an aperture in the condenser plane. Different groups [92, 93, 94] have produced holograms using slightly different recipes.

The general formula for the phase is

$$\alpha(\rho) = \frac{2\pi}{\lambda}\left(-\Delta f \rho^2 + \frac{1}{4}C_S \rho^4\right) + g\rho \cos(\theta), \tag{57}$$

The inline approach is recovered when $g = 0$. For correction to be applicable over a wide field (beyond a standard STEM probe), it is necessary to have a phase ranging over $4 - 6\pi$. One can use either a continuous slope with thickness $t = \alpha$ or a discontinuous slope $t = Mod(\alpha, 2\pi)$ with $2\pi$ phase wraps. The first approach results in a thick membrane and significant absorption, while the second approach requires precise tuning of the discontinuities.

An inline version with a large value of $\Delta f$ can be used to create many beams that are in focus at different values of the *z* coordinate [95]. Although any kind of groove, such as a sinusoid, can be used, this approach has not been used so far (see Fig. 32b). Here, we describe the off-axis approach, which allows excellent control of the phase by employing a thin membrane at the cost of spurious diffraction orders. Grillo *et al.* [94] explained how to remove such spurious orders by the smart use of optics components. Figure 32a shows a typical aberration function in the presence of defocus ($C_S$ = 0.5 mm and $\Delta f = 40$ nm). The corresponding off-axis phase S-CGH and a realization are shown in Figs 32c and 32d, respectively.

For realization of the hologram, it is preferable to use a sinusoidal or a blazed groove shape, which ensures a smoother variation of the phase, as described in section 1.6. For practical reasons, the carrier frequency must be quite large so that isolating a specific beam in the diffraction plane (with the desired $C_S$ value) is easier. These constraints naturally lead to the use of very large holograms, resulting in challenges in fabrication and durability, as mentioned in section 2.5.

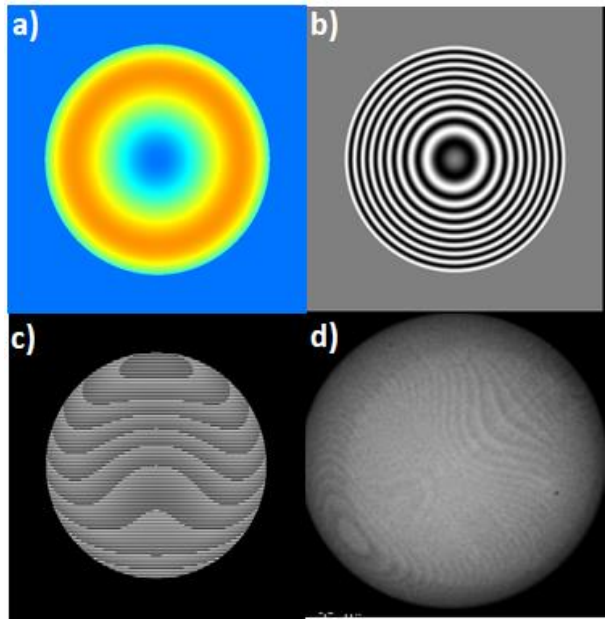

*Figure 32: Design of a holographic $C_S$ corrector. a) Desired phase plate; b) Inline S-CGH with a sinusoidal groove for Δf = 400 nm; c) Off-axis S-CGH; d) Realization in silicon nitride.*

**Chapter 4 – Conclusions**

In this tutorial, we have reviewed the concept of "imaging" holography and explored synthetic holography, from the use of CGHs to simulate interference patterns to the possibility of engineering wave functions and new techniques in materials science. We have provided mathematical descriptions of the most commonly used groove profiles in amplitude and phase S-CGHs, discussing their efficiency are and the design of mixed phase and amplitude S-CGHs. We have described two fabrication techniques that can be used to manufacture S-CGHs, including their optimization and limitations. Finally, we have provided examples of possible uses of S-CGHs in the field of electron vortex beams. We have highlighted the fact that real phase S-CGHs are sensitive to imperfections introduced by fabrication, reducing their efficiency.

**Acknowledgments**

This project has received funding from the European Union's Horizon 2020 research and innovation programme (Grant No. 766970, project "Q-SORT"; Grant No. 856538, project "3D MAGiC"; Grant No. 823717, project "ESTEEM3"), from the Deutsche Forschungsgemeinschaft (Project-ID 405553726 – TRR

270), from the DARPA TEE program (Grant MIPR# HR0011831554) and from an Ontario Early Researcher Award (ERA) and a Canada Research Chair (CRC).

**Credits**

The following article has been submitted to/accepted by Journal of Applied Physics. After it is published, the link at which it will be found at will be provided.

**Data Availability Statement**

The data that support the findings of this study are available from the corresponding author upon reasonable request.

*Appendices*

*Appendix A1:*

This appendix contains in-depth calculations of different profiles for phase S-CGHs.

*Cosine profile*

A sinusoidal/ cosinusoidal profile can be described by the periodic function $f(\alpha) = \frac{1}{2}(1 + \cos(\alpha(\vec{\rho})))$, with a transmission function of the form

$$T(\vec{\rho}) = e^{i\frac{\tilde{a}}{2}\cos(\alpha(\vec{\rho}))} e^{i\frac{\tilde{a}}{2}} =$$
$$= e^{i\frac{a_1}{2}\cos(\alpha(\vec{\rho}))} e^{-\frac{a_2}{2}\cos(\alpha(\vec{\rho}))} e^{i\frac{\tilde{a}}{2}}$$
$$= e^{ia_1'\cos(\alpha(\vec{\rho}))} e^{-a_2'\cos(\alpha(\vec{\rho}))} e^{i\tilde{a}'} \,, \quad (A1)$$

where the primed variables are used to simplify the calculations. $T(\vec{\rho})$ can be rewritten using the Jacobi-Anger expansion

$$T(\vec{\rho}) = e^{i\tilde{a}'} \sum_{n=-\infty}^{\infty} \left(i^n J_n(\tilde{a}')\right) e^{in\alpha(\vec{\rho})} \,, \quad (A2)$$

Where the Fourier series coefficients of $T(\vec{\rho})$ are

$$\tau_n = i^n J_n(\tilde{a}') e^{i\tilde{a}'} \,. \quad (A3)$$

$|\tau_n|^2$ can be plotted with the complex argument Bessel Function, while an analytical approximation can be derived from Eq. A1. The first term of $T(\vec{\rho})$ can be rewritten using the Jacobi-Anger expansion, while the second term can be expanded by making use of the approximation $a_2' \ll 1$, resulting in the expression

$$T(\rho) \approx \exp(ia) \exp(ia_1 \cos(\alpha))(1 - a_2 \cos(\alpha)) \,. \quad (A4)$$

Then:

$$T(\rho) \approx \exp(i\tilde{a}') \left[\sum_n (-i)^n J_n(a'_1) \exp(in\alpha)\right] (1 - a'_2 \cos(\alpha)) .. \quad (A5)$$

If the Fourier coefficient is defined according to the expression

$$\tau_m = \int T(\rho) \exp(-im\alpha)\, d\alpha \quad (A6)$$

and we use the property of convolution

$$\tau_m = \exp(i\tilde{a}')\left\{\int\left[\sum_n (-i)^n J_n(a'_1)\exp(in\alpha)\right]\exp(-im\alpha)\,d\alpha\right\} * \left\{\int (1 - a'_2 \cos(\alpha))\exp(-im\alpha)\,d\alpha\right\}, \quad (A7)$$

then we can develop the terms

$$\int\left[\sum_n (-i)^n J_n(a'_1)\exp(in\alpha)\right]\exp(-im\alpha)\,d\alpha = \sum_n \delta_{m,n} J_n(a'_1)(-i)^n \quad (A8)$$

$$\int (1 - a_2 \cos(\alpha))\exp(-im\alpha)\,d\alpha = \int \left(1 - \frac{a'_2}{2}(\exp(i\alpha) + \exp(-i\alpha))\right)\exp(-im\alpha)\,d\alpha =$$

$$= \delta_{m,0} - \frac{a'_2}{2}(\delta_{m,1} + \delta_{m,-1}) \quad (A9)$$

to obtain the final coefficient in the form

$$\tau_m = (-i)^m \big(J_m(a'_1)\big) * \left[\delta_{m,0} - \frac{a'_2}{2}(\delta_{m,1} + \delta_{m,-1})\right]\exp(i\tilde{a}'). \quad (A10)$$

If we now make use of discrete convolution according to the expression

$$\tau_m = \exp(i\tilde{a}')\sum_k (-i)^k \big(J_k(a'_1)\big)\left[\delta_{k,m-0} - \frac{a'_2}{2}(\delta_{k,m-1} + \delta_{k,m+1})\right], \quad (A11)$$

then

$$\tau_m = (-i)^m J_m(a_1) - \frac{a_2}{2}((-i)^{m+1} J_{m+1}(a'_1) + (-i)^{m-1} J_{m-1}(a_1))\exp(i\tilde{a}') \quad (A12)$$

$$\tau_m = (-i)^m \{J_m(a_1) + \frac{ia_2}{2}[J_{m-1}(a_1) - J_{m+1}(a_1)]\}\exp(i\tilde{a}')$$

The efficiency of the $n^{th}$ diffraction order is proportional to $|\tau_n|^2$, where

$$|\tau_m|^2 = \left[J_m^2(a'_1) - \frac{{a'_2}^2}{4}(J_{m-1}(a'_1) - J_{m+1}(a'_1))^2\right]e^{-2a'_2} . \qquad (A13)$$

*Squared profile*

A periodic grating function with a squared profile $f(\alpha) = \frac{1}{2}Sign|sin(\alpha(\vec{\rho}))|$ has the transmission function

$$T(\vec{\rho}) = e^{i\frac{\tilde{a}}{2}Sign(sin(\alpha(\vec{\rho})))} =$$
$$= e^{i\tilde{a}'Sign(sin(\alpha(\vec{\rho})))}$$
$$= e^{ia'_1 Sign(sin(\alpha(\vec{\rho})))} \cdot e^{-a'_2 Sign(sin(\alpha(\vec{\rho})))} , \qquad (A14)$$

where the $\tilde{a}$ is primed to take into account that the amplitude of $Sign(sin(\alpha(\vec{\rho})))$ is half the peak-to-valley distance, simplifying the calculation. The Fourier coefficients of $T(\vec{\rho})$ can be calculated from the expression

$$\tau_n = \frac{1}{2\pi}\int_0^{2\pi} e^{i\tilde{a}'Sign(sin(\alpha(\vec{\rho})))} e^{-in\alpha(\vec{\rho})} d\alpha . \qquad (A15)$$

As a result of the properties of the *Sign* function, Eq. A15 can be written

$$\tau_n = \frac{1}{2\pi}\left[\int_0^{\pi} e^{i\tilde{a}'} e^{-in\alpha(\vec{\rho})} d\alpha + \int_{\pi}^{2\pi} e^{-i\tilde{a}'} e^{-in\alpha(\vec{\rho})} d\alpha\right]$$
$$= \begin{cases} \cos(\tilde{a}') & \text{for } n = 0 \\ 0 & \text{for } n \text{ even} \\ \dfrac{2\sin(\tilde{a}')}{n\pi} & \text{for } n \text{ odd} , \end{cases} \qquad (A16)$$

such that, for example,

$$\tau_1 = \frac{2\sin(\tilde{a}')}{\pi} =$$
$$= \frac{2}{\pi}[\sin(a'_1)\cosh(a'_2) + i\cos(a'_1)\sinh(a'_2)] . \qquad (A17)$$

The efficiency of the first diffracted order is proportional to

$$|\tau_1|^2 = \frac{4}{\pi^2}[\sin^2(a'_1)\cosh^2(a'_2) + \cos^2(a'_1)\sinh^2(a'_2)] \qquad (A18)$$

The maximum is reached when $a'_1 \sim 1.57\ rad$, so the optimal peak-to-valley phase difference is $\Delta\varphi \sim \pi$.

*Triangular profile*

We now consider a triangularly-shaped profile, specifically an isosceles triangle that can be described by the profile function $f(\alpha) = \frac{1}{\pi}(Sign(sin(\alpha(\vec{\rho}))))(\pi - Mod(\alpha(\vec{\rho}), 2\pi))$, with the transmission function

$$T(\vec{\rho}) = e^{i\tilde{a}\frac{1}{\pi}\left(Sign(sin(\alpha(\vec{\rho})))\right)(\pi - Mod(\alpha(\vec{\rho}), 2\pi))}, \qquad (A19)$$

where $Mod(a, b)$ is the remainder after dividing $a$ by $b$. The Fourier coefficients can be calculated from the integral

$$\tau_n = \frac{1}{2\pi}\int_0^{2\pi} e^{i\tilde{a}\frac{1}{\pi}\left(Sign(sin(\alpha(\vec{\rho})))\right)(\pi - Mod(\alpha(\vec{\rho}), 2\pi))} e^{-in\alpha(\vec{\rho})} d\alpha, \qquad (A20)$$

resulting in the expression

$$\tau_n = \frac{-i(a_1 + ia_2)\left[(-1)^{n+1} + e^{i(a_1 + ia_2)}\right]}{(a_1^2 + 2ia_1 a_2 - a_2^2 - n^2\pi^2)}, \qquad (A21)$$

from which the generic efficiency of the $n^{th}$ diffracted order is proportional to

$$|\tau_n|^2 = \frac{(a_1^2 + a_2^2)[1 + 2(-1)^{n+1} e^{-a_2}(\cos(a_1)) + e^{-2a_2}]}{[a_1^4 + 2a_1^2 a_2^2 + a_2^4 + n^4\pi^4 - 2n^2 a_1^2\pi^2 + 2n^2 a_2^2\pi^2]}. \qquad (A22)$$

*Blazed profile*

A specific example of a triangular profile is a blazed profile, which is similar to that of a sawtooth blade. The profile function is now simply $f(\alpha) = \frac{1}{2\pi}\left(Mod(\alpha(\vec{\rho}), 2\pi)\right)$, while the transmittance function is

$$T(\vec{\rho}) = e^{i\tilde{a}\frac{1}{2\pi}(Mod(\alpha(\vec{\rho}), 2\pi))}. \qquad (A23)$$

As before, the Fourier coefficients are

$$\tau_n = \frac{1}{2\pi} \int_0^{2\pi} e^{i\tilde{a}\frac{1}{2\pi}(Mod(\alpha(\vec{\rho}),2\pi))} e^{in\alpha(\vec{\rho})} d\alpha =$$

$$= \frac{-i(-1+e^{i\tilde{a}})}{(\tilde{a}+2\pi n)} \, , \tag{A24}$$

such that

$$|\tau_n|^2 = \frac{(1+e^{-2a_2} - 2\cos(a_1)e^{-a_2})}{[(a_1+2\pi n)^2 + a_2^2]} \tag{A25}$$

***Appendix A2:*** Calculations for amplitude S-CGHs.

For most grating profiles, the calculations are relatively simple as they just involve calculating the squared modulus of the Fourier coefficients of the profile functions. However, some cases require full calculations.

***Squared profile with an arbitrary duty cycle***

For a square profile with an arbitrary duty cycle, the profile function between 0 and $2\pi$ is

$$f(\alpha) = \begin{cases} 1 & for \ 0 < \alpha \leq D*2\pi \\ 0 & for \ D*2\pi < \alpha \leq 2\pi \end{cases}, \tag{A26}$$

where the duty cycle $D$ is constant, with $0 < D < 1$. The fundamental frequency associated with $f(\alpha)$ is $\omega_0 = 1$ since the period is $2\pi$. The Fourier coefficients are

$$\tau_0 = \frac{1}{2\pi} \int_0^{2\pi} f(\alpha) d\alpha = \frac{1}{2\pi} \int_0^{D*2\pi} 1 \, d\alpha = D \tag{A27}$$

$$\tau_{n\neq 0} = \frac{1}{2\pi} \int_0^{2\pi} f(\alpha) e^{in\omega_0 \alpha} d\alpha = \frac{1}{2\pi} \int_0^{D*2\pi} 1 e^{in\alpha} d\alpha =$$

$$= \frac{1}{2\pi} \frac{1}{in} \left[ e^{inD2\pi} - 1 \right] = \frac{1}{2} \frac{e^{inD\pi}}{in\pi} \left[ e^{inD\pi} - e^{-inD\pi} \right] =$$

$$= \frac{e^{inD\pi}}{n\pi} \sin(nD\pi) = De^{inD\pi} sinc(nD\pi) \, , \tag{A28}$$

Therefore:

$$|\tau_n|^2 = \begin{cases} D^2 & for\ n = 0 \\ D^2 sinc^2(nD\pi) & for\ n \neq 0 \end{cases}. \qquad (A29)$$

# References

1. D. Gabor, Nature, 161, 777 (1948).

2. A. Claverie, Transmission Electron Microscopy in Micro-nanoelectronics. Hoboken, New York, Wiley, 2013.

3. M. Takeda, H. Ina, and S. Kobayashi, J. O.S.A, 72, 156 (1982).

4. L.B.Lesem, P.M.Hirsch, J.A.Jordan. Jr, Sci.Applications, 19 (1968).

5. V.Grillo, E.Rotunno, Ultramicroscopy, 125, 97 (2012).

6. V.Y. Bazhenov, M.V. Vasnetsov, and M.S.Soskin, Jetp Lett, 429 (1990).

7. N.R. Heckenberg, R. McDuff, C.P. Smith, H. Rubinsztein-Dunlop, and M.J.Wegener, Opt. Quantum Electronics, 24, S951-S962 (1992).

8. E. Bolduc, N. Bent, E. Santamato, E.Karimi, and R.W.Boyd, Optics Letters, Vol. 38, 3546 (2013).

9. A. H. Tavabi, P. Rosi, E. Rotunno, A. Roncaglia, L. Belsito, S. Frabboni, G. Pozzi, G. C. Gazzadi, P.-H. Lu, R. Nijland, M. Ghosh, P. Tiemeijer, E. Karimi, R. E. Dunin-Borkowski, and V. Grillo, Phys. Rev. Lett., 126, 094802 (2021).

10. R. F. Egerton, Electron Energy-Loss Spectroscopy in the Electron Microscope, 3rd ed. New York, Springer, (2011).

11. K. Iakoubovskii, K. Mitsuishi, Y. Nakayama and K. Furuya, Physical Review B, 77, 1 (2008).

12. G. Pozzi, Parcticles and waves in electron optics and microscopy. In Peter W Hawkes, Advances in Imaging and Electron Physics. New York, Elsavier Academic Press, (2016).

13. H.Reimer, H.Kohl. Transmission Electron Microscopy, 3rd ed. Springer Series in Optical Sciences. New York, Springer, (1993), Vol. 36.

14. V.Grillo, G.C.Gazzadi, E.Karimi, E.Mafakheri, R.W. Boyd and S.Frabboni, Applied Physics Letters, 104, 1, (2014).

15. T. R. Harvey, J. S. Pierce, A. K. Agrawal, P. Ercius, M. Linck and B. J. McMorran, New Journal of Physics, 14, 093039 (2014).

16. S.Bhattacharyya, C.T. Koch and M. Rühle, Ultramicroscopy, 106, 525 (2006).

17. Roy Shiloh, Yossi Lereah, Yigal Lilach, Ady Arie, Ultramicroscopy, 144, 26 (2014).

18. L. Grünewald, D. Gerthsen and S. Hettler, Beilstein J. Nanotechnol, 10, 1290 (2019).

19 A. Auslender, M. Halabi, G. Levi, O. Diéguez, A. Kohn, Ultramicroscopy, 198, 18 (2019).

20. M. Schowalter, J. T. Titantah, D. Lamoen, and P. Kruse, Appl. Phys. Lett., 89, 112102 (2005).

21. M. Wanner, D. Bach, D. Gerthsen, R. Werner, B. Tesche, Ultramicroscopy, 106, 341 (2006).

22. A. Harscher, H. Lichte. Proc Int Conf Electron Microsc ICEM14 (1), 553 (1998).

23. A.Sanchez, M.A.Ochando, J.Phys. C: Solid State Phys, 18, 33 (1985).

24. D. B. Williams, C. Barry Carter. Transmission Electron Microscopy - A textbook for Marterial Science. New York, Springer (2009).

25. V. Grillo, E. Karimi, R. Balboni, G. C. Gazzadi, F. Venturi, S. Frabboni, J. S Pierce, B. J. McMorran and R. W Boyd, Microscopy and Microanalisis, 21, 503 (2015).

26. B.J. McMorran, A. Agrawal, I.M. Anderson, A.A. Herzing, H.J. Lezec, J.J. McClelland, J. Unguris, Science, 331, 192 (2011).

27. W-H. Lee, Applied Optics, 18 (1979).

28. W.H. Lee, Applied Optics, 13, 1677 (1974)

29. L.A.Giannuzzi, F.A.Stevie. Introduction to Focused Ion Beam. New York, Springer (2005).

30. T. Schachinger, A. Steiger-Thirsfeld, S. Löffler, M. Stöger-Pollach, S. Schneider, D. Pohl, B. Rellinghaus, P. Schattschneider, European Microscopy Congress 2016: Proceedings, 717 (2016)

31. E. Mafakheri, A.H.Tavabi, P.-H. Lu, S.Frabboni, V.Grillo, Applied Physics Letters, 110, 1 (2017).

32. W. W. Hu, K. Sarveswaran, M. Lieberman and G. H. Bernstein, Journal of Vacuum Science & Technology B: Microelectronics and Nanometer Structures Processing, Measurement, and Phenomena, 22 (2004).

33. M. Häffner, A. Haug, A. Heeren, M. Fleischer, H. Peisert, T. Chassé, and D. P. Kern, Journal of Vacuum Science & Technology B: Microelectronics and Nanometer Structures Processing, Measurement and Phenomena, 25 (2007).

34. S. Hettler, L. Radtke, L. Grünewald, Y. Lisunova, O. Peric, J. Brugger, S. Bonanni, Micron., 127, 102753 (2019).

35. L. Clark, A. Béché, G. Guzzinati, A. Lubk, M. Mazilu, R. Van Boxem, and J. Verbeeck, Phys. Rev. Lett., 111, 064801 (2013).

36. J. Verbeeck, A. Béché, K. Müller-Caspary, G. Guzzinati, M. A. Luong, M. D. Hertog, Ultramicroscopy, 190, 58 (2018).

37. B. J. McMorran, A. Agrawal, P. A. Ercius, V. Grillo, A. A. Herzing, T. R. Harvey, M. Linck, J. S. Pierce, Phil. Trans. R. Soc. A, 375 (2016).

38. M. Uchida, A. Tonomura, Nature, 464, 737 (2010).

39. J. Verbeeck, H. Tian & P. Schattschneider, Nature, 467, 301 (2010).

40. G. Pozzi, P-H. Lu, A. H. Tavabi, M. Duchamp, R. E.Dunin-Borkowski, Ultramicroscopy, 181, 191 (2017).

41. A. H. Tavabi, H. Larocque, P.-H. Lu, M. Duchamp, V. Grillo, E. Karimi, R. E. Dunin-Borkowski and G. Pozzi, Phys. Rev. Research, 2, 013185 (2020).

42. A. Béché, R. Van Boxem, G. Van Tendeloo and J. Verbeeck, Nature Phys, 10, 26 (2014).

43. K. Y. Bliokh, I. P. Ivanov, G. Guzzinati, L. Clark, R. Van Boxem, A. Béché, R. Juchtmans, M. A. Alonso, P. Schattschneider, F. Nori, J. Verbeeck, Phys. Rep., 690, 1 (2017).

44. R. Shiloh, P-H. Lu, R. Remez, A. H. Tavabi, G. Pozzi, R. E. Dunin-Borkowski and A. Arie, Physica Scripta, 94 (2019).

45. J. Harris, V. Grillo, E. Mafakheri, G.C. Gazzadi, S. Frabboni, R. W. Boyd & E. Karimi, Nature Physics, 11, 629 (2015)

46. H.Larocque, I. Kaminer, V. Grillo, G. Leuchs, M.J. Padgett, R.W. Boyd, M. Segev, E. Karimi, Contemporary Physiscs, 59 (2018)

47. A.Béché, R.Winkler, H.Plank, F.Hofer, J.Verbeeck, Micron, 80, 34 ( 2016).

48. F. Venturi, New approaches for phase manipulation and characterisation in the transmission electron microscope. (2018).

49. C. W. Johnson, D. H. Bauer and B. J. McMorran, Applied Optics., 59, 1594 (2020).

50. E. Karimi, G. Zito, B. Piccirillo, L. Marrucci and E. Santamato, Optics Letters, 32, 3053 (2007)

51. R. Guenther, Modern Optics. New York, Wiley, (1996).

52. A. M. Yao, M. J. Padgett, Adv. in Opt. and Phot, 3 (2011).

53. L.G.Gouy, Comptes Rendus de l'Académie des Sciences Paris, 110, (1890).

54. S.Feng, H.G.Winful, Opt. Lett., 26, 485 ( 2001).

55. T.D.Visser, E.Wolf, Opt. Commun., 283, 3371 (2010).

56. S.M.Loyd, M.Babiker, G. Thirunavukkarasu and J.Yuan, Reviews of Modern Physics, 89, 035004 (2017).

57. M. Abramowitz, I.A. Stegun, Handbook of mathematical functions, 10th ed., U.S. Government Printing Office, (1972).

58. K. Y. Bliokh, P. Schattschneider, J. Verbeeck, F. Nori, Phys. Rev. X, 2 (2012).

59. P. Schattschneider, T. Schachinger, M. Stoeger-Pollach, S. Loeffler, A. Steiger-Thirsfeld, K. Y. Bliokh, and F. Nori, Nature Commun., 5 (2014).

60. F. Venturi, M. Campanini, G. C. Gazzadi, R. Balboni, S. Frabboni, R. W. Boyd, R. E. Dunin-Borkowski, E. Karimi and V. Grillo, Applied Physics Letters, 111 (2017).

61. F. Venturi, R. Balboni, G. C. Gazzadi, M. Campanini, E. Karimi, V. Grillo, S. Frabboni, R. W. Boyd, European Microscopy Congress 2016: Proceedings, (2016).

62. C. W. Johnson, J. S. Pierce, R. C. Moraski, A. E. Turner, A. T. Greenberg, W. S. Parker, B. J. McMorran, Optics express, 28 (2020).

63. L.Allen, M.W. Beijersbergen, R.J.C. Spreeuw and J.P,Woerdman, Phys.Rev.A, 45 (1992).

64. P. Schattschneider, M. Stoeger-Pollach, J. Verbeeck, Phys. Rev. Lett., 109 (2012).

65. V. Grillo, A. H. Tavabi, F. Venturi, H. Larocque, R. Balboni, G.C. Gazzadi, S. Frabboni, P-H. Lu, E. Mafakheri, F. Bouchard, R. Dunin-Borkowski, R. W. Boyd, M. Lavery, M. Padgett and E. Karimi, Nature Communications, 8 (2017).

66. E. Rotunno, M. Zanfrognini, S. Frabboni, J. Rusz, R. E. Dunin-Borkowski, E. Karimi and V. Grillo, Phys. Rev. B, 100 (2019).

67. M. Zanfrognini, E. Rotunno, S. Frabboni, A. Sit, E. Karimi, U. Hohenester and V. Grillo, ACS Photonics, 6, 620 (2019).

68. J.Durnin, J. Opt. Soc. Am. A, 4 (1987).

69. J. Durnin, J. J. Miceli and J. H. Eberly, Phys. Rev. Lett., 58 (1987).

70. J. Durnin, J. J. Miceli and J. H. Eberly, Phys. Rev. Lett., 66 (1991).

71. D. McGloin, K. Dholakia, Contemporary Physics, 46 (2005).

72. Y. Lin, W. Seka, J. H. Eberly, H. Huang and D. L. Brown, Applied Optics, 31, 2708 (1992).

73. J. H. McLeod, Journal of the Optical Society of America, 44 (1954).

74. A. Thaning, Z. Jaroszewicz and A. T. Friberg, Applied Optics, 42 (2003).

75. Z. Bin, L. Zhu, Applied Optics, 37 (1998).

76. T. Tanaka, S. Yamamoto, Optics Communications, 184, 113 (2000).

77. A. Vasara, J. Turunen and A. T. Friberg, Journal of the Optical Society of America A, 6 (1989).

78. J. A. Davis, E. Carcole and D. M. Cottrell, Applied Optics, 35, 593 (1996).

79. J. A. Davis, E. Carcole and Don M. Cottrell. 4, 1996, Applied Optics, Vol. 35, pp. 599-602.

80. K. Uehara, H. Kikuchi, Applied Physics B, 48, 125 (1989).

81. A. J. Cox and D. C.Dibble, Journal of the Optical Society of America A, 9, 282 (1992).

82. V. Grillo, E. Karimi, G. C. Gazzadi, S. Frabboni, M. R. Dennis and R. W. Boyd, Physical Review X, 4 (2014).

83. V. Grillo, J.Harris, G.C.Gazzadi, R.Balboni, E.Mafakheri, M.R.Dennis, S.Frabboni, R.W. Boyd, E.Karimi, , Ultramicroscopy, 166, 48 (2016).

84. K. Saitoh, K. Hirakawa, H. Nambu, N. Tanaka and M. Uchida, Journal of the Physical Society of Japan, 85 (2016).

85. C. Zheng, T. C. Petersen, H. Kirmse, W. Neumann, M. J. Morgan and J. Etheridge, Physical Review Letters, 119 (2017).

86. P. A. Midgley and R. E. Dunin-Borkowski, Nature Materials, 8, 271 (2009).

87. G. Guzzinati, W. Ghielens, C. Mahr, A. Béché, A. Rosenauer, T. Calders and J. Verbeeck, Applied Physics Letters, 114 (2019).

88. S. Hettler, L. Grünewald and M. Malac, New Journal of Physics, 21 (2019).

89. E. Rotunno, A.H. Tavabi, E. Yucelen, S. Frabboni, R.E. Dunin Borkowski, E. Karimi, B.J. McMorran, and V. Grillo, Phys. Rev. Applied, 11, 044072 (2019).

90. M. Haider, H. Rose, S. Uhlemann, B. Kabius, K. Urban, Journal of Electron Microscopy, 47, 395 (1998).

91. O. L. Krivanek, N. Dellby, A. R. Lupini, Ultramicroscopy, 78, 1 (1999).

92. R. Shiloh, R. Remez, P.-H. Lu, L. Jin, Y. Lereah, A. H. Tavabi, R. E. Dunin-Borkowski, A. Arie, Ultramicroscopy, 189, 46 (2018).

93. M. Linck, P. A. Ercius, J. S. Pierce, B. J. McMorran, Ultramicroscopy, 182, 36 (2017).

94. V. Grillo, A. H. Tavabi, E. Yucelen, P.-H. Lu, F. Venturi, H. Larocque, L. Jin, A. Savenko, G. C. Gazzadi, R. Balboni, S. Frabboni, P. Tiemeijer, R. E. Dunin-Borkowski and E. Karimi, Optics Express, 25, 21851 (2017).

95. J.Verbeeck, H.Tian, A. Béché, Ultramicroscopy, 113, 83 (2012).